\documentclass[fleqn,usenatbib]{mnras}
\pdfoutput=1
\usepackage{newtxtext,newtxmath}
\usepackage[T1]{fontenc}
\usepackage{romannum}
\usepackage{graphicx} 
\usepackage[tight,footnotesize]{subfigure}
\usepackage{amsmath}	
\usepackage{multirow}
\usepackage{comment}
\usepackage{textcomp}
\usepackage[center]{caption}
\usepackage{hyperref}
\hypersetup{
    colorlinks=true,
    linkcolor=blue,
    filecolor=magenta,      
    urlcolor=blue,
    citecolor=blue,
}
\newcommand{\so}{${\omega}_{\rm o}$\,}
\newcommand{\pso}{P$_{\omega_{\rm o}}$\,}
\newcommand{\pom}{P$_{\omega_{\rm -}}$\,}
\newcommand{\pop}{P$_{\omega_{\rm +}}$\,}
\newcommand{\ptwo}{P$_{2\omega_{\rm o} + \rm N}$\,}
\newcommand{\psotwo}{P$_{2\omega_{\rm o}}$\,}
\newcommand{\pthree}{P$_{3\omega_{\rm o} + \rm 2N}$\,}
\newcommand{\pf}{P$_{\omega_{\rm o} + \rm 2N}$\,}
\newcommand{\pomtwo}{P$_{2\omega_{\rm -}}$\,}
\newcommand{\pomthree}{P$_{3\omega_{\rm -}}$\,}

\newcommand{\orb}{${\Omega}$\,} 
\newcommand{\twoorb}{${2\Omega}$\,} 
\newcommand{\s}{${\omega}$\,} 
\newcommand{\be}{$\omega$\,-\,$\Omega$\,} 
\newcommand{\twobe}{$2(\omega$\,-\,$\Omega)$\,} 
\newcommand{\threebe}{3($\omega$\,-\,$\Omega$\,)} 
\newcommand{\beplus}{$\omega$\,+\,$\Omega$\,} 
\newcommand{\sone}{$2\omega$\,-\,$\Omega$\,} 
\newcommand{\stwo}{$2\omega$\,-\,$3\Omega$\,} 
\newcommand{\sthree}{$\omega$\,-\,$2\Omega$\,} 
\newcommand{\po}{P$_{\Omega}$\,} 
\newcommand{\ptwoo}{P$_{2\Omega}$\,} 
\newcommand{\pthreeo}{P$_{3\Omega}$\,} 
\newcommand{\pseveno}{P$_{7\Omega}$\,} 
\newcommand{\peighto}{P$_{8\Omega}$\,} 
\newcommand{\pnineo}{P$_{9\Omega}$\,} 
\newcommand{\pteno}{P$_{10\Omega}$\,} 
\newcommand{\peleveno}{P$_{11\Omega}$\,} 
\newcommand{\ps}{P$_{\omega}$\,} 
\newcommand{\pb}{P$_{\omega-\Omega}$\,} 
\newcommand{\pbplus}{P$_{\omega+\Omega}$} 
\newcommand{\ptwob}{P$_{2(\omega-\Omega)}$} 
\newcommand{\pthreeb}{P$_{3(\omega-\Omega)}$} 
\newcommand{\psone}{P$_{2\omega-\Omega}$} 
\newcommand{\pstwo}{P$_{2\omega-3\Omega}$} 
\newcommand{\psthree}{P$_{\omega-2\Omega}$} 

\title[Unveiling the nature of LS Cam, V902 Mon, and J0746]{A Step towards Unveiling the Nature of Three Cataclysmic Variables: LS Cam, V902 Mon, and SWIFT J0746.3-1608}
\author[Rawat et al.]
{
Nikita Rawat, $^{1,2}$\thanks{E-mail: nikita@aries.res.in (NR), jeewan@aries.res.in (JCP)}
J. C. Pandey, $^{1}$ \footnotemark[1]
Arti Joshi, $^{3}$ 
Umesh Yadava $^{2}$
\\
$^{1}$Aryabhatta Research Institute of Observational sciencES (ARIES), Nainital 263001, India\\
$^{2}$Deen Dayal Upadhyaya Gorakhpur University, Gorakhpur 273009, India\\
$^{3}$School of Physics and Technology, Wuhan University, Wuhan 430072, China
}

\date{Accepted XXX. Received YYY; in original form ZZZ}

\pubyear{2021}

\begin{document}
\label{firstpage}
\pagerange{\pageref{firstpage}--\pageref{lastpage}}
\maketitle
\begin{abstract}
We have carried out detailed time-resolved timing analyses of three cataclysmic variables (CVs) namely LS Cam, V902 Mon, and SWIFT J0746.3-1608, using the long-baseline, high-cadence optical photometric data from the Transiting Exoplanet Survey Satellite (TESS). Our analysis of LS Cam observations hints the presence of a superorbital period of $\sim$ 4.025$\pm$0.007 d along with negative and positive superhump periods of $\sim$ 3.30 h and 3.70 h, respectively. These results can be explained as an interaction of nodal and apsidal precession of the accretion disc with orbital motion. For the other two sources, V902 Mon and SWIFT J0746.3-1608, we have found evidence of a beat period of 2387.0$\pm$0.6 s and 2409.5$\pm$0.7 s, respectively, which were not found in earlier studies. Our results presented in this study indicate the change in the accretion mode during the entire observing period for both sources. For V902 Mon, an apparent orbital period derivative of (6.09 $\pm$ 0.60) $\times 10^{-10}$ was also found. Moreover, the second harmonic of orbital frequency dominates the power spectrum of SWIFT J0746.3-1608, suggestive of ellipsoidal modulation of the secondary star. Present analyses suggest that LS Cam could be a superhumping CV whereas V902 Mon and SWIFT J0746.3-1608 are likely to be variable disc-overflow accreting intermediate polars.
\end{abstract}


\section{Introduction}
Cataclysmic Variables (CVs) are the semi-detached binary systems in which primary is a magnetic white dwarf (WD) that accretes material through a Roche-lobe filling late-type main-sequence star, also known as a secondary star. The magnetic field of the WD plays a crucial role in governing the accretion process in these binaries. It also decides two distinct classes of magnetic CVs (MCVs): polars and intermediate polars (IPs). In polars, the magnetic field of the white dwarf is strong enough (typically, in a range of 10-100 MG) to lock the whole system into synchronous (or almost synchronous) rotation. It also prevents the formation of an accretion disc and the accreting material channels directly to the WD magnetic pole(s). While in the case of IPs, magnetic field of the WD is weaker (typically, 1-10 MG) and an accretion disc can form which is disrupted at the magnetospheric radius. Hence, the material from the secondary is accreted either through an accretion stream or an accretion disc or a combination of both. The majority of IPs have spin period of the WD (\ps), roughly the one-tenth of the orbital period (\po) of the binary system (\ps $\sim$ 0.1 \po) 
and orbital periods longer than the `period gap' of 2-3 h \citep{2010MNRAS.401.2207S}. Generally, the higher values of X-ray luminosities of IPs than those of polars is attributed to the higher accretion rate.

\par The high energy X-rays originate from the accretion shock and optical emission is seen due to the reprocessing of these X-rays in the surface layers of the disc (including the hotspot) and/or the atmosphere of the secondary. The WD rotation modulates emission in the X-ray and optical regions due to the obscuration by the intervening accretion curtains and reprocessed radiation from axisymmetric parts of the accretion disc. Reprocessing of the beam from parts of the system that rotate with the binary orbital frequency, such as an inflated part of the disc (the hotspot) or the secondary itself, produces an optical beat pulsation. Further, obscurations of the WD by the material rotating in the binary frame and eclipse of the hotspot by an optically thick disc (or ring) give rise to the orbital modulation \citep[][and references therein]{1986MNRAS.219..347W}. 

\begin{table*}
\centering
\caption{Observation log of sources where start and end time are in calendar date.}
\label{tab:obslog}
\renewcommand{\arraystretch}{1.4}
\begin{tabular}{lcccc}
\hline
Source Name & Sector & Start time & End time & Total observing days\\
\hline
\multirow{4}{*}{LS Cam} & 19 & 2019-11-28T14:04:20.676 & 2019-12-23T15:26:46.080 & 24.8\\
& 20 & 2019-12-25TT00:08:45.948 & 2020-01-20T07:48:03.141 & 26.3\\
& 26 & 2020-06-09T18:17:54.123 & 2020-07-04T15:07:55.466 & 24.9\\
& 40 & 2021-06-25T03:33:33.297 & 2021-07-23T08:24:18.088 & 28.2\\
\hline
V902 Mon & 33 & 2020-12-18T05:40:40.133 & 2021-01-13T01:50:30.998 & 25.8 \\
\hline
SWIFT J0746.3-1608 & 34 & 2021-01-14T06:29:41.446 & 2021-02-08T13:39:28.238 & 25.0\\
\hline
\end{tabular}
\end{table*}

\par There are primarily three accepted scenarios for accretion in IPs and the feasibility of each one of them depends on the magnetic field strength of WD and mass accretion rate. The first is the disc-fed accretion, in which an accretion disc is present in the system, which is disrupted at the magnetosphere radius. From this radius, material flows along the magnetic field lines resulting in the formation of `accretion curtains' near the magnetic poles of the WD \citep{1988MNRAS.231..549R}. The second is the disc-less or stream-fed accretion, in which the high magnetic field of the WD does not allow the formation of a disc and infalling material is channelised along the magnetic field lines to the pole caps \citep{1986MNRAS.218..695H}. In the third possibility, known as disc-overflow accretion \citep{1989ApJ...340.1064L, 1996ApJ...470.1024A}, disc-fed and stream-fed accretions can simultaneously occur as a part of the accretion stream skims over the disc and then interacts with the magnetosphere of the WD \citep{1989MNRAS.238.1107H, 1991ApJ...378..674K}. In all these accretion scenarios, whenever accretion takes place via a disc, the accreting material impacts on to both magnetic poles and whenever accretion takes place via a stream, accretion occurs at both poles continuously with varying accretion rate and only a fraction of accretion flow flips between two poles.
\par A featured characteristic of IPs is the presence of multiple periodicities in the X-ray and optical power spectra because of the complex interactions between the spin and orbital modulations. Therefore, the presence of spin, beat, orbital, and other sideband frequencies in the power spectra and their amplitudes play a vital tool in distinguishing the mode of accretion in the system. In the disc-fed accretion, modulation at the spin frequency (\s) of the white dwarf occurs \citep{1995A&A...298..165K, 1996MNRAS.280..937N}, whereas stream-fed accretion gives rise to modulation  at the lower orbital sideband of the spin frequency, i.e. beat  (\be) frequency \citep{1991MNRAS.251..693H, 1992MNRAS.255...83W}. \cite{1992MNRAS.255...83W} also showed that if there is an asymmetry between the magnetic poles, stream-fed accretion can also produce a modulation at the spin frequency, in addition to that at the beat frequency. Hence, \sone  frequency plays an important role in distinguishing between these two modes of accretion in X-ray bands and only present in disc-less systems along with sometimes dominant \orb component, \be, and \s. For a disc-overflow accretion, where disc-fed and stream-fed simultaneously occur, modulations at both \s and \be  frequencies are expected to occur \citep[see][]{1991MNRAS.251..693H, 1993MNRAS.265L..35H}. Further, the variable nature of accretion flow is one of the basic characteristics of an IP and has been observed in IPs e.g. TX Col and FO Aqr for a number of times \citep[see][for details]{1992MNRAS.254..705N, 1993MNRAS.265L..35H, 1996rftu.proc..123B, 1997MNRAS.289..362N, 1999ASPC..157...47W, 2020ApJ...896..116L, 2021ApJ...912...78R}. 

In this paper, we present a detailed investigation of three CVs namely LS Cam, V902 Mon, and SWIFT J0746.3-1608, based on their high cadence long baseline TESS observations. These sources are taken from the intermediate polar catalogue of Koji Mukai\footnote{\url{http://asd.gsfc.nasa.gov/Koji.Mukai/iphome/iphome.html}}. Our aim is to explore the true nature of all three sources.

\par The paper is structured as follows. Section \ref{sec2} reviews each of the sources individually and in section \ref{sec3}, we describe observations and data used for this study. Section \ref{sec4}, \ref{sec5}, and \ref{sec6} contains our analysis and results for LS Cam, V902 Mon, and J0746, respectively. Finally, discussion and summary are presented in Sections \ref{sec7} and \ref{sec8}, respectively.

\begin{figure*}
\centering
\subfigure[]{\includegraphics[width=16cm, height=5cm]{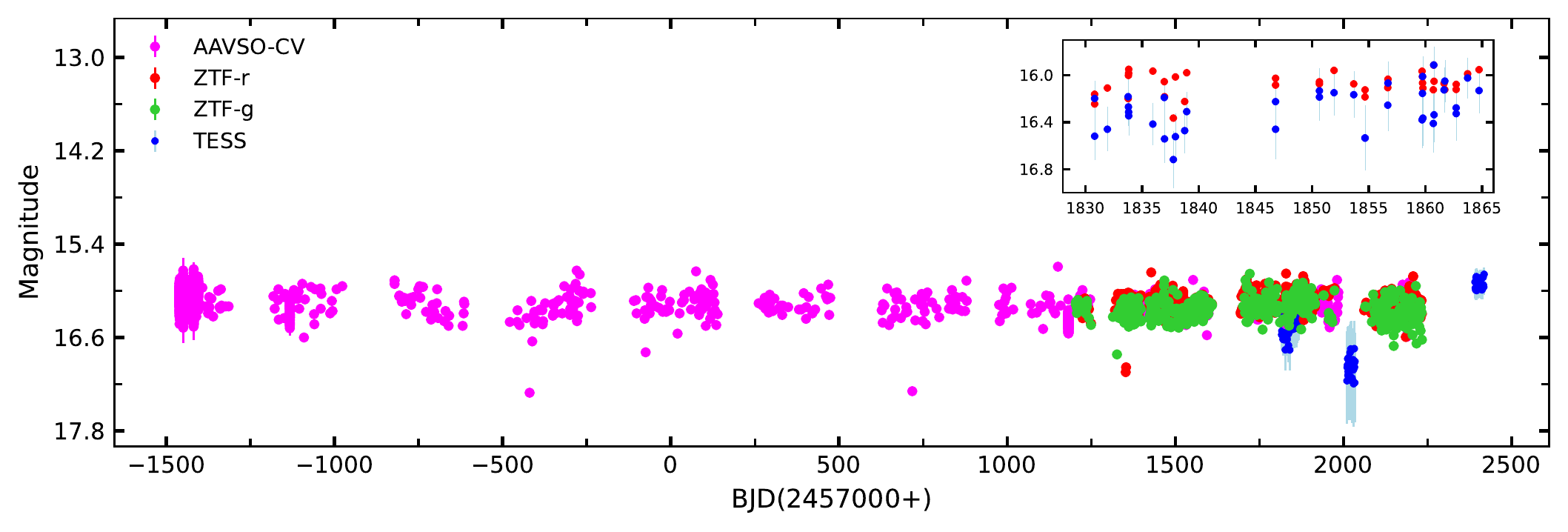} \label{fig:lc-lscam-all}}
\subfigure[]{\includegraphics[width=16cm, height=5cm]{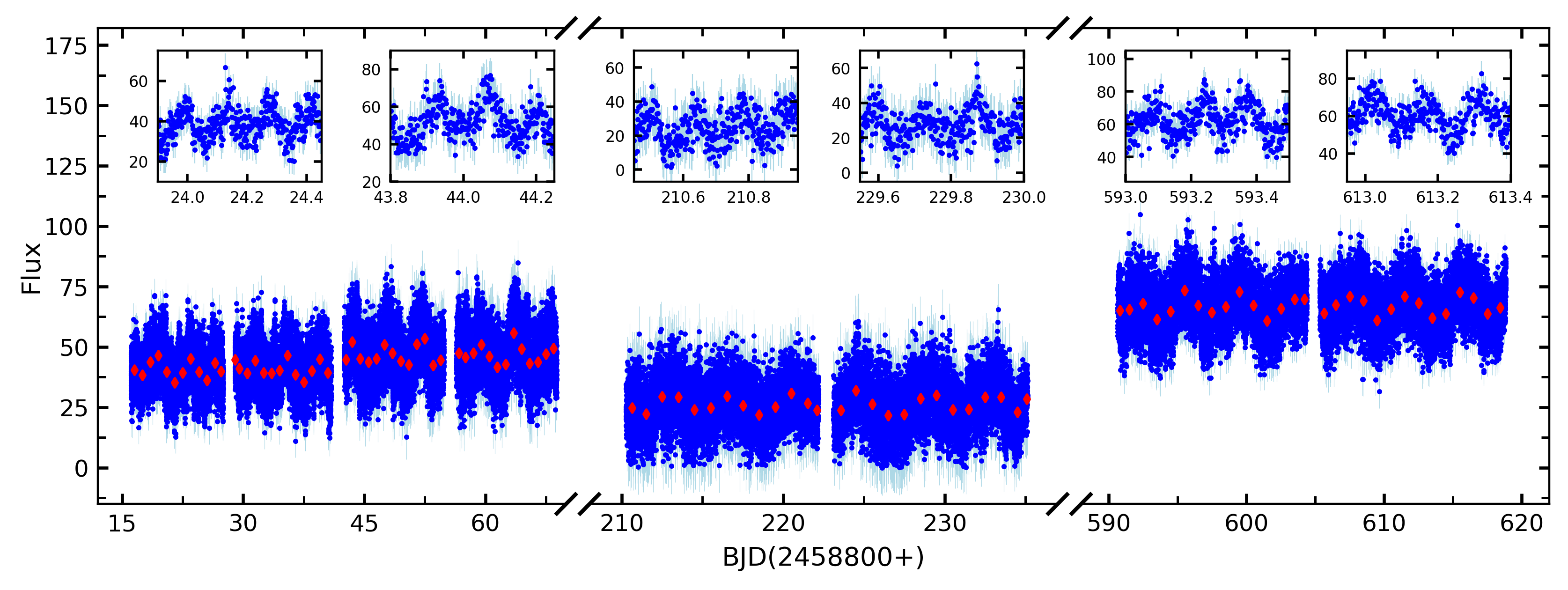} 
\label{fig:lc-lscam}}
\caption{(a) Long-term variable light curve of LS Cam as observed from AAVSO, ZTF, and TESS. Inset of the figure shows the simultaneous observations from TESS and ZTF-r. (b) TESS light curve of LS Cam of the sectors 19, 20, 26, and 40, where red diamonds represent the mean flux of each day. Inset of the figure shows closeup of some variability cycles.}
\end{figure*}

\section{Review of Sources}\label{sec2}
\subsection{LS Cam}
LS Cam (HS 0551+7241) was discovered as a CV by \cite{1998AJ....115.1634D} using spectroscopic observations. It is located at a distance of  $1.7_{-0.1}^{+0.1}$ kpc \citep{2021A&A...650C...3G}. The identification spectrum of LS Cam, obtained in 1995, revealed features typical for a CV: H \Romannum{1}, He \Romannum{2}, and He \Romannum{1} emission lines on the top of a blue continuum. The periodogram analysis of the H$\rm \alpha$, H$\rm \beta$, and He \Romannum{2} $ \rm \lambda$4686 radial velocities revealed that the hydrogen lines are likely dominated by the S-wave component, following the orbital variations with a period of 4 h. At the same time, He \Romannum{2} $\lambda$4686 line shows rapid fluctuations of $\sim$ 50 minutes. So, based on these variations, \cite{1998AJ....115.1634D} concluded that LS Cam might be an IP. Additionally, from the analysis of emission-line radial velocity curves, \citet{2017RNAAS...1...29T} confirmed the orbital period of LS Cam to be 3.417 h.

\subsection{V902 Mon}
The source V902 Mon was identified  as a CV in the Isaac Newton Telescope (INT) Photometric H$\alpha$ Survey of the northern Galactic plane (IPHAS) survey due to its prominent H$\alpha$ emission \citep{2007MNRAS.382.1158W}. Using the optical photometric observations, \citet{2012ApJ...758...79A} classified it as a deeply eclipsing IP  with an orbital period of 8.162 h and spin period of WD of  $\sim$ 2210 s. Further, the spin modulation was found to be varying in their observing runs between 2006 and 2009, sometimes too weak to be noticeable. However, between 2008 and 2017, \cite{2018A&A...617A..52W} found the spin period of 2208 s at multiple epochs and concluded that V902 Mon accretes via disc only due to the absence of beat modulations. The distance of V902 Mon is found to be $3.1_{-0.5}^{+0.8}$ kpc \citep{2021A&A...650C...3G}.

\subsection{SWIFT J0746.3-1608}
SWIFT J0746.3-1608 (hereafter J0746) was discovered in the Swift/BAT survey and found to be a highly variable X-ray source at a distance of $625_{-8}^{+10}$ pc \citep{2021A&A...650C...3G}. From optical observations, \cite{2013AJ....146..107T} found an orbital period of 9.38 h and reported that J0746 might belong to the class of either Nova-like variable or an IP. \cite{2019MNRAS.484..101B} concluded that J0746 might be an IP with a plausible spin period of $\sim$ 2300 s. However, due to the short X-ray coverage, they could not separate this periodicity from the beat and other sidebands most commonly found in the power spectra of IPs. From comparative studies of XMM-Newton observations of 2016 and 2018, they also confirmed that J0746 has returned to its high state in 2018.\\

\begin{figure*}
\includegraphics[width=15cm, height=9.5cm]{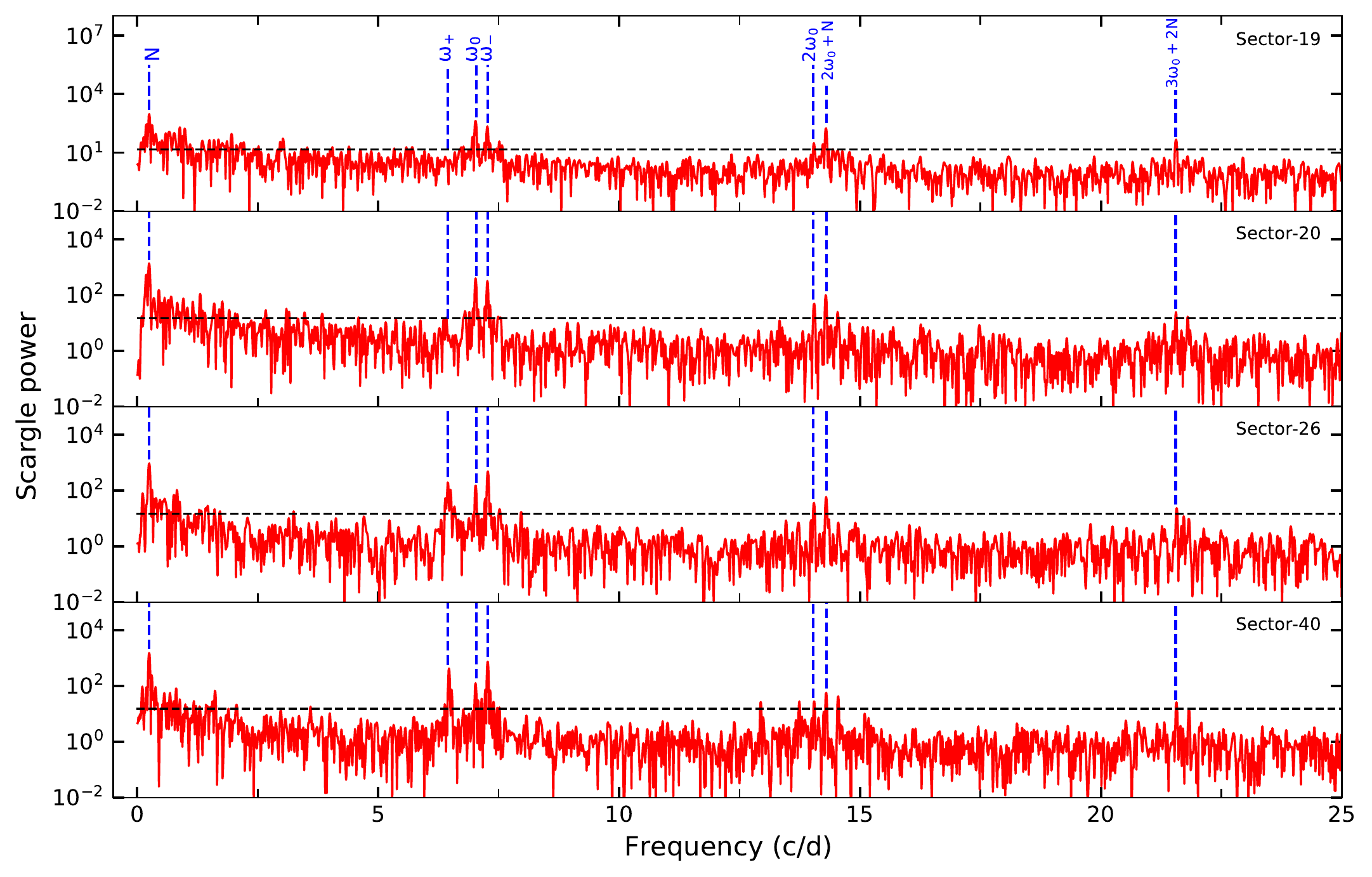}
\caption{Power spectra of LS Cam of all sectors. Major frequencies are marked for clear visual inspection.}
\label{fig:ps-lscam}
\end{figure*}

\section{Observations and Data}
\label{sec3}
We have used the archival data from TESS for all three sources. The TESS instrument consists of four wide-field CCD cameras, each with field-of-view of 24\textdegree $\times$ 24\textdegree so that all cameras can image a region of the sky measuring 24\textdegree $\times$ 96\textdegree. TESS observations are broken up into sectors, each lasting two orbits, or about 27.4 days and conducts its downlink of data while at perigee. This results in a small gap in the data compared to the overall run length. The TESS bandpass extends from 600-1000 nm with an effective wavelength of 800 nm \cite[see][for details]{2015JATIS...1a4003R}. The log of observations for each source is given in Table \ref{tab:obslog}. The data was stored in Mikulski Archive for Space Telescopes data archive\footnote{\url{https://mast.stsci.edu/portal/Mashup/Clients/Mast/Portal.html}} with unique identification numbers `TIC 138042556', `TIC 334061567', and `TIC 234712743', for LS Cam, V902 Mon, and J0746, respectively. The cadence for each source was 2 minutes; however, for J0746 the data was available at a cadence of 20 seconds also. For our analysis, we have taken `PDCSAP' flux, which is the simple aperture photometry (SAP) flux after removing the common instrumental systematics from it using the co-trending basis vectors. The PDCSAP flux also corrects for the amount of flux captured by the photometric aperture and crowding from known nearby stars\footnote{See Sec. 2.0 of the TESS Archive Manual at \url{https://outerspace.stsci.edu/display/TESS/2.0+-+Data+Product+Overview}}. The data taken during an anomalous event had quality flags greater than 0 in the FITS file data structure, and thus we have considered only the data with the `quality flag' = 0.\\

Wherever available, we have also used ZTF\footnote{\url{https://www.ztf.caltech.edu/}}-r, ZTF-g, AAVSO\footnote{\url{https://www.aavso.org/}}-CV (Clear or unfiltered reduced to V sequence), AAVSO-V (Johnson V), and AAVSO-I (Johnson I) to represent the long-term behaviour of sources. The ground-based data is available only for the source LS Cam and V902 Mon. We found a positive linear dependence of TESS magnitude on the simultaneous ground-based magnitudes, which indicates that PDCSAP flux values appear to be following the flux observed from ground-based observations.  Nonetheless,  as noted in the TESS archival manual\footnote{See Sec. 2.1 of the TESS Archive Manual at \url{https://outerspace.stsci.edu/display/TESS/2.1+Levels+of+data+processing}} and also pointed out by \citet{2021AJ....162...49L} that in the PDCSAP data, the periodic variability should be unaffected but slow, aperiodic variability (such as that from a low-amplitude outburst) might be removed. Thus, the results presented in this manuscript can be taken with caution if aperiodic variabilities such as quasi-periodic oscillations have been removed from PDCSAP flux values after processing from SAP flux values.

\section{LS Cam} 
\label{sec4}
\subsection{Light Curve and Power Spectral Analysis} \label{ls}

Figure \ref{fig:lc-lscam-all} shows the long-term light curve of LS Cam, where the variable nature of the source is visible. The average change in magnitude in the AAVSO-CV band is between $\sim$ 16.6 - 15.7 and on some occasions, it has gone to a fainter state with a magnitude more than 17.3. To overplot the TESS light curve, we have converted TESS flux unit of electrons per second to magnitude using the zero-point value of 20.44 \citep{2018.V, 2021ApJ...908...51F}. The inset of Figure \ref{fig:lc-lscam-all} shows some of the simultaneous observations between ZTF-r and TESS, which confirms that during these observations, LS Cam was always in a relatively stable high state. Figure \ref{fig:lc-lscam} shows the TESS light curve of LS Cam. There is a gap of almost five months between the observations in sectors 20 and 26 and a gap of one year between sector 26 and 40. Further, sector 26 has a mean flux value of 26.22 $\pm$ 0.08 $e^{-}/s$ compared to the mean flux value of 40.70 $\pm$ 0.05 $e^{-}/s$, 46.92 $\pm$ 0.07 $e^{-}/s$, and 66.87 $\pm$ 0.05 $e^{-}/s$ of the sectors 19, 20, and 40, respectively.  Unfortunately, there were no simultaneous observations corresponding to sector 26 observations, but similar low states were  also seen during the previous observations corresponding to BJDs 2456580.39, 2457718.29, and 2458353.99. This may indicate that that there might have been a change in the system's state during the sector 26 observations.

\par To find the periodic signals in the data, we have performed the Lomb-Scargle (LS) periodogram analysis \citep{1976Ap&SS..39..447L, 1982ApJ...263..835S}. The LS power spectrum of all four sectors is shown in Figure \ref{fig:ps-lscam}. Several peaks are present in the power spectrum of LS Cam. The significance of these detected peaks is determined by calculating the false-alarm probability (FAP; \cite{1986ApJ...302..757H}). The horizontal dashed line in each power spectrum represents the 90\% confidence level. The derived periods from the periodogram analyses are given in Table \ref{tab:periods-lscam}.  The periods corresponding to four dominant peaks from the combined data of all sectors are 4.025 $\pm$ 0.007 d, 3.4171 $\pm$ 0.0002 h, 3.3007 $\pm$ 0.0002 h, and 1.67896 $\pm$ 0.00005 h. 

\begin{figure*}
\centering
\includegraphics[width=6.7cm, height=5.5cm]{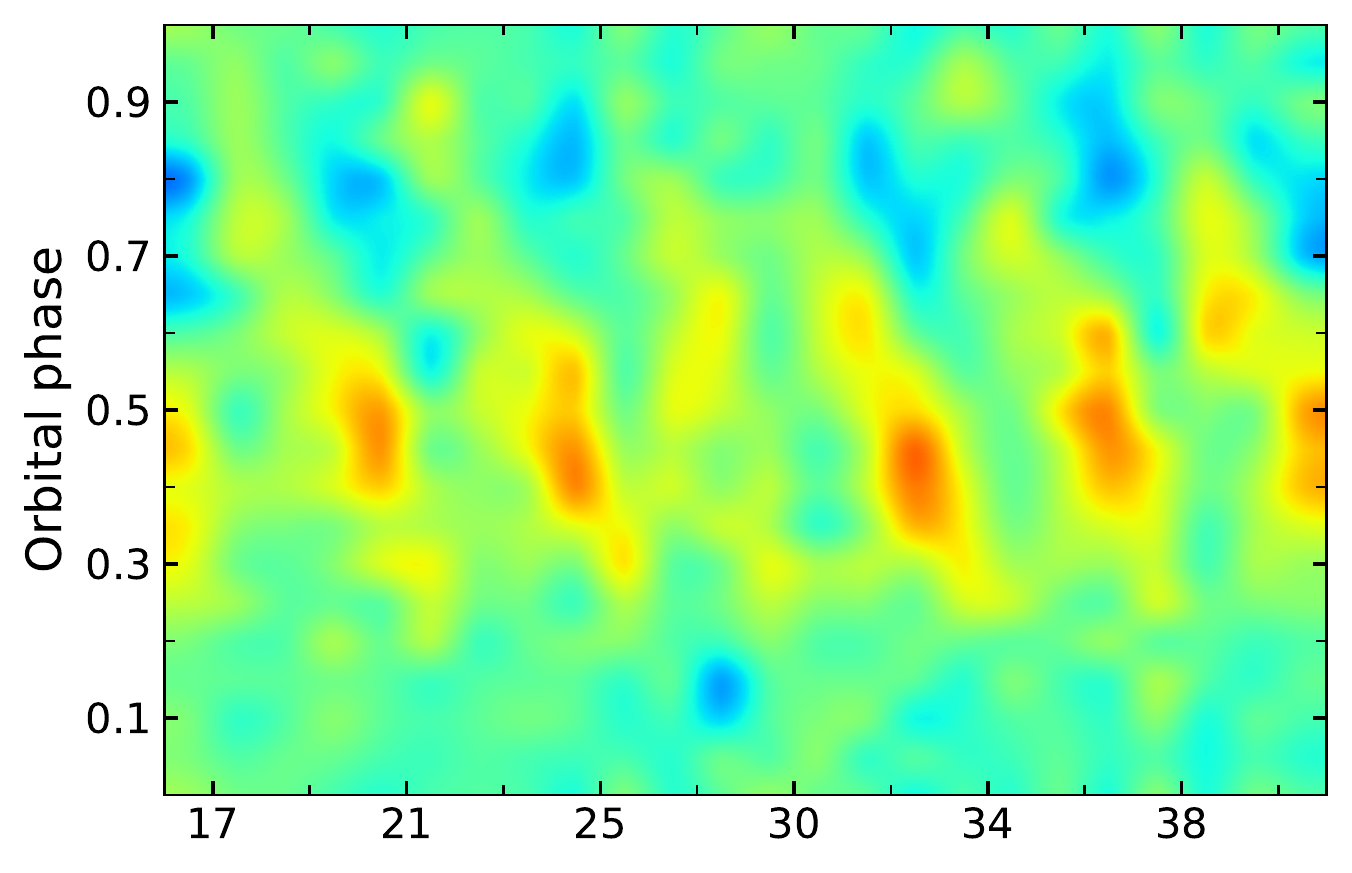}
\hspace{0.5cm}
\includegraphics[width=7.7cm, height=5.6cm]{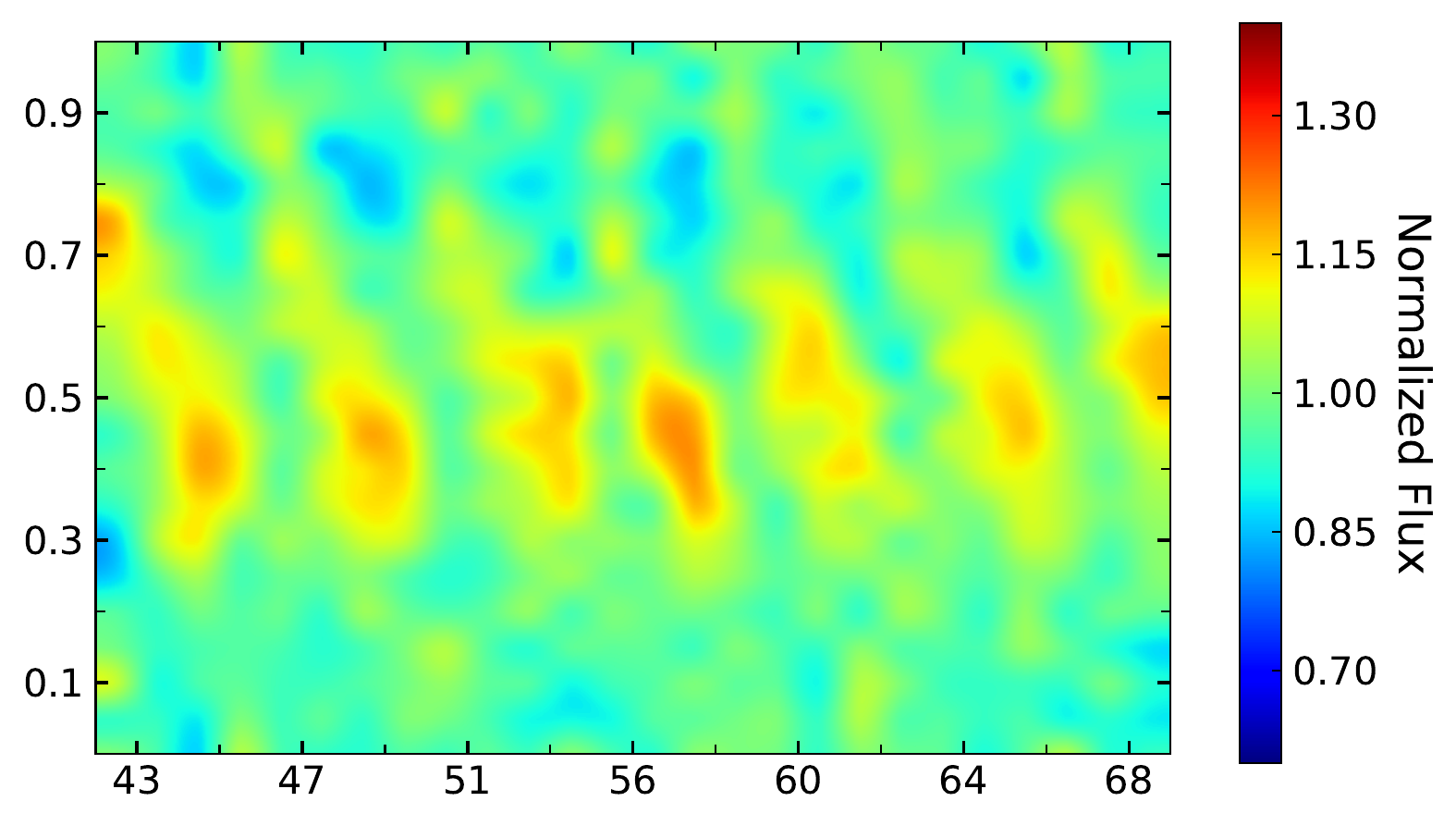}
\includegraphics[width=6.7cm, height=5.5cm]{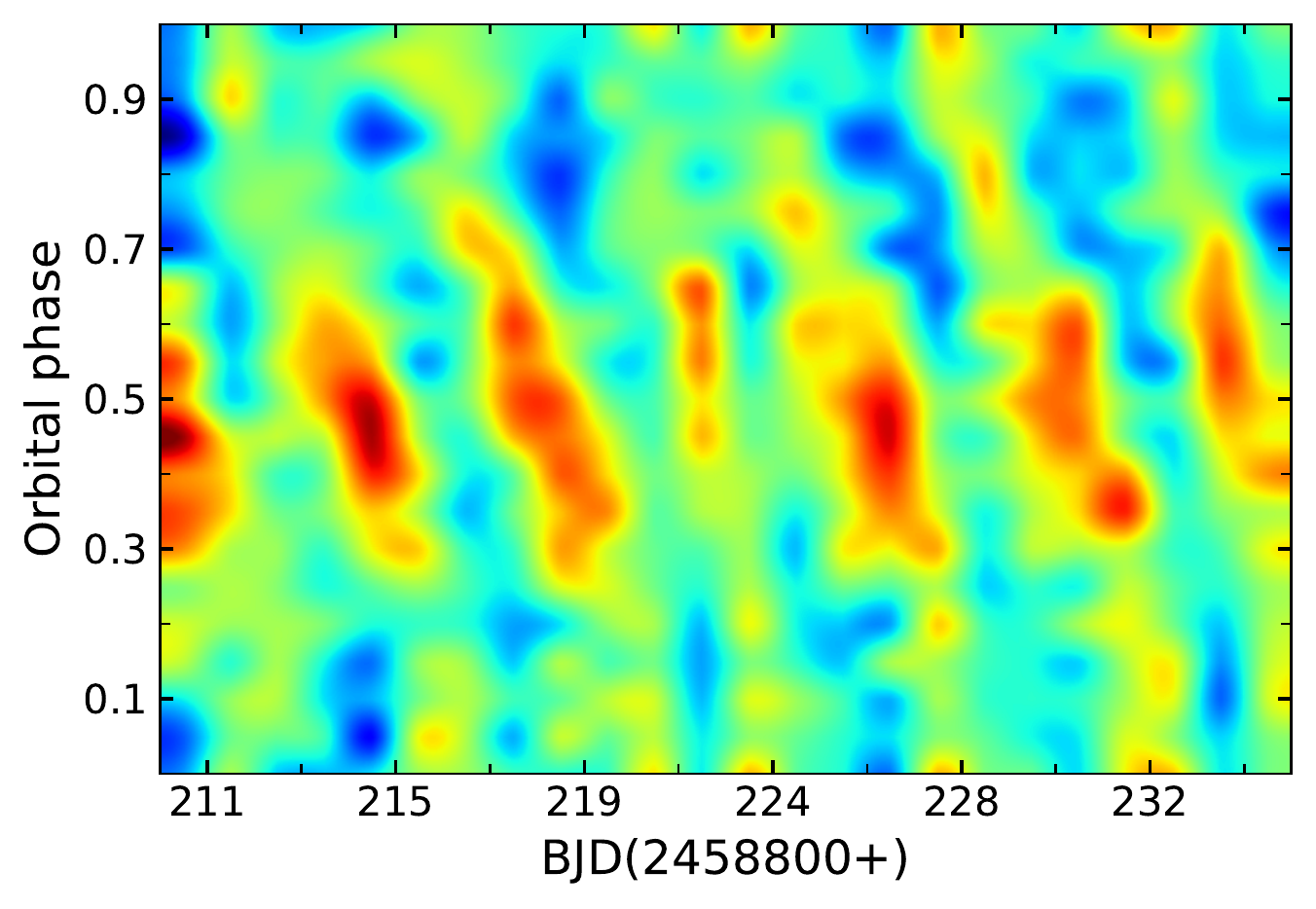}
\hspace{0.5cm}
\includegraphics[width=7.7cm, height=5.6cm]{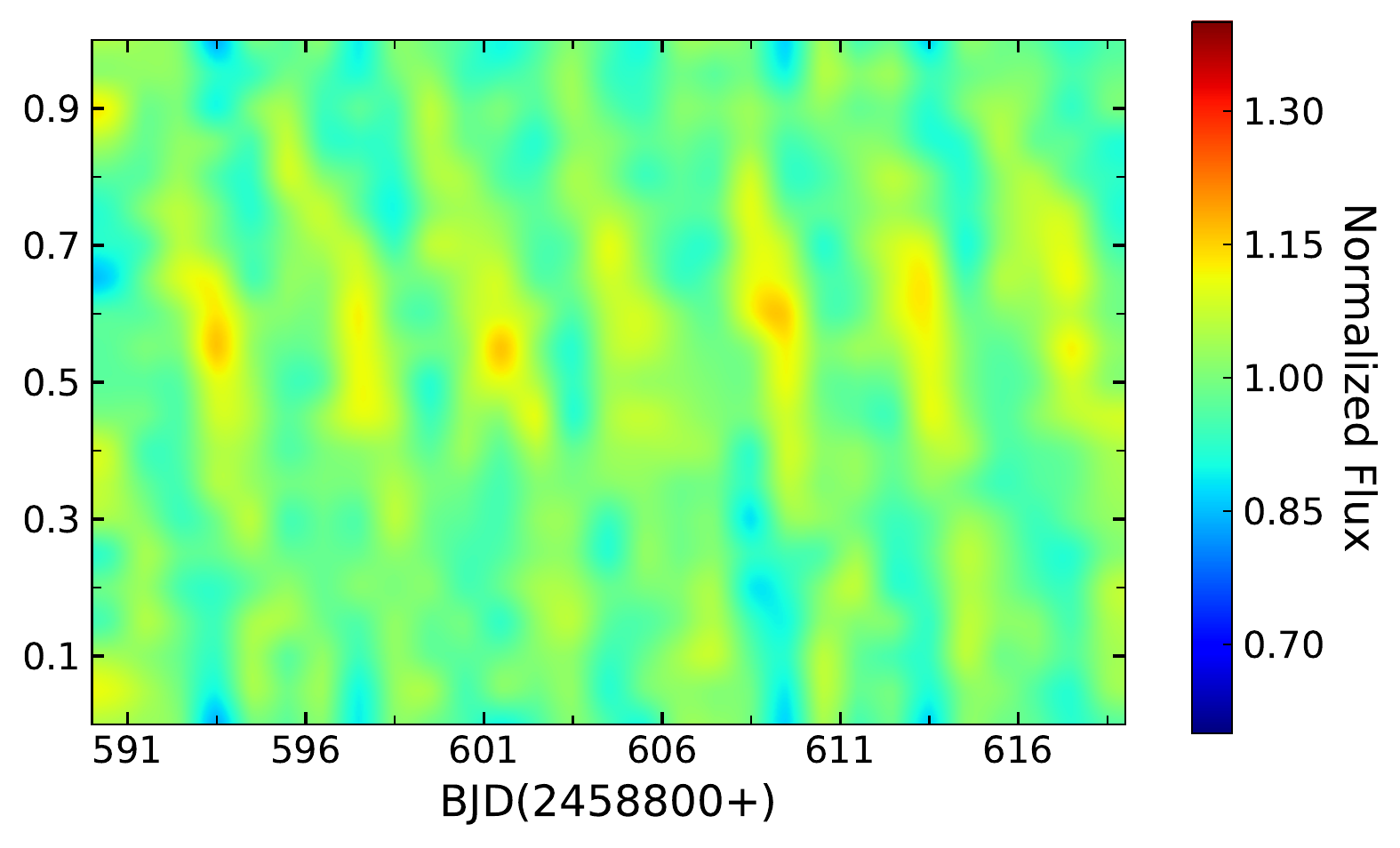}
\caption{Phase folded light curves of LS Cam for orbital period. The top two  panels correspond to sectors 19 and 20, whereas the bottom two panels correspond to sectors 26 and 40, respectively. The colourbars (normalized flux) have been used for representing the modulations.}
\label{fig:orb-lscam}
\end{figure*}

The 3.417 h periodicity is identified as the orbital period of the system. 
For the orbital phase folding the data, we have taken the one-day time-resolved data segments to see the short-term variations. The reference time for folding was taken to be the observation starting time as extremums could not be adequately identiﬁed in the light curve. Each light curve was folded with the binning of 20 points in a phase. We have shown the orbital phase folded profile in Figure \ref{fig:orb-lscam}. The top 2 panels correspond to the observations of sectors 19 and 20, whereas the bottom two panels correspond to the observations of sectors 26 and 40, respectively. The orbital modulation was found to be highest in sector 26, lowest in sector 40, and intermediate in sectors 19 and 20. In all sectors, a $\sim$ 4-day periodic variation in all orbital phase folded profiles seems to be taking place.

Other frequencies identified in power spectral analyses can be explained  based on the following two scenarios. 

\subsubsection{Scenario A} \label{scenarioA}

In this scenario, if we consider 1.67896 h as spin period (\ps) of WD, the beat period (\( P_{\omega-\Omega}  = \frac{P_{\Omega} P_\omega}{P_{\Omega} - P_{\omega}}\)) is estimated to be 3.30 h, which is present in all four sectors' observation.  
Considering these values of orbital, spin, and beat periods, the periods corresponding to various harmonics \sthree, \twoorb, \twobe, \threebe, \sone, and \stwo should be 4.025 d, 1.71 h, 1.65 h, 1.10 h, 1.11h, and 3.19 h, respectively, which are present in the power spectrum of LS Cam. The presence of all these frequencies speculates that LS Cam can be an IP. However, as \beplus was not found to be present in the power spectrum, therefore the maximum value of the amplitude of \sthree allowed by the model for optical power spectra of IPs predicted by \citet{1986MNRAS.219..347W} should be half of that of \be, which was also found for V1223 Sgr by \citet{1984MNRAS.206..261W}. For LS Cam, \sthree was the largest-amplitude signal in all sectors, which weakens its possibility of being an IP based on the above-identified frequencies. Further, if considered all frequencies belong to an IP class for LS Cam then  we have no explanation for the origin of $\sim$ 3.7 h periodicity in sectors 26 and 40.

\begin{table*}
\begin{center}
\caption{Periods corresponding to the dominant frequency peaks in the power spectra of LS Cam of the observations of all four sectors.}
\label{tab:periods-lscam}
\end{center}
\renewcommand{\arraystretch}{1.4}

\begin{tabular}{lccccc}
\hline
\multirow{2}{*}{Identification} & \multicolumn{4}{c}{Period (days$^{*}$/hours)}\\
\cline{2-6}
 & Sector 19 & Sector 20 & Sector 26 & Sector 40 & Combined\\ 
 \hline
 P$_{\rm N}$$^{\dagger}$/\psthree${^\ddagger}$ & 3.97 $\pm$ 0.16$^{*}$ & 4.05 $\pm$ 0.16$^{*}$ & 3.98 $\pm$ 0.16$^{*}$ & 4.03 $\pm$ 0.14$^{*}$ & 4.025 $\pm$ 0.007$^{*}$\\
\pso${^\dagger}$/\po${^\ddagger}$ & 3.416 $\pm$ 0.005 & 3.419 $\pm$ 0.005 & 3.415 $\pm$ 0.005 & 3.418 $\pm$ 0.004 & 3.4171 $\pm$ 0.0002 \\
\pom${^\dagger}$/\pb${^\ddagger}$ & 3.302 $\pm$ 0.004 & 3.303 $\pm$ 0.004 & 3.297 $\pm$ 0.004 & 3.302 $\pm$ 0.004 & 3.3007 $\pm$ 0.0002\\
\pop${^\dagger}$ &  ------ & ------ & 3.724 $\pm$ 0.006 & 3.709 $\pm$ 0.005 & ------ 
\\
\ptwo${^\dagger}$/\ps${^\ddagger}$ & 1.679 $\pm$ 0.001 & 1.680 $\pm$ 0.001 & 1.679 $\pm$ 0.001 & 1.678 $\pm$ 0.001 & 1.67896 $\pm$ 0.00005\\
\psotwo${^\dagger}$/\ptwoo${^\ddagger}$ & 1.709 $\pm$ 0.001 & 1.708 $\pm$ 0.001 & 1.709 $\pm$ 0.001 & 1.708 $\pm$ 0.001 & 1.70860 $\pm$  0.00005\\
\pthree${^\dagger}$/\psone${^\ddagger}$ & 1.1132 $\pm$ 0.0005 & 1.1135 $\pm$ 0.0005 & 1.1129 $\pm$ 0.0005 & 1.1132 $\pm$ 0.0005 & 1.11316 $\pm$ 0.00002\\
\pf${^\dagger}$/\pstwo${^\ddagger}$ & 3.192 $\pm$ 0.004 & 3.202 $\pm$ 0.004 & 3.191 $\pm$ 0.004 & 3.193 $\pm$ 0.004 & ------ \\
\pomtwo${^\dagger}$/\ptwob${^\ddagger}$ & ------ & 1.651 $\pm$ 0.001 & ------ & 1.650 $\pm$ 0.001 & ------\\
\pomthree${^\dagger}$/\pthreeb${^\ddagger}$ & ------ & 1.1009 $\pm$ 0.0005 & ------ & 1.0996 $\pm$ 0.0004 & ------\\
\hline
\end{tabular}

{\textbf{Notes:}} {
$\dagger$: Identification of periods based on scenario B,
$\ddagger$: Identification of periods based on scenario A. Please see Sections \ref{scenarioA} and \ref{scenarioB} for details.\\}
\end{table*}

\subsubsection{Scenario B} \label{scenarioB}
Considering the orbital period (\pso) of the system as 3.41 h then the periodicity of 3.30 h can be a negative superhump (\pom) because it has slightly less period than the \pso. Thus the 4.025 d period could be the superorbital period (P$_{\rm N}$). In this assumption, the periods of 3.19 h, 1.71 h, 1.68 h, 1.65 h, 1.11 h, 1.10 h can be designated as \pf, \psotwo,  \ptwo, \pomtwo, \pthree, and \pomthree,  respectively. 
There are mainly two possibilities for the presence of several days superorbital period in CVs. The first one is due to the ``short outbursts'' of ER UMa-type dwarf novae \citep[see][for details]{1995PASP..107..443R}. Since no outbursts have been detected for LS Cam, therefore the superorbital period found here can not be attributed to the outbursts of ER UMa-type dwarf novae. However, superorbital period in CVs is also associated with retrograde precession of an accretion disc \citep[see][for details]{1997PASP..109.1100P}. Generally, these systems satisfy the relation \so + N = $\omega_{-}$, where $\omega_{-}$ corresponds to the frequency of a ``negative superhump''. In this way, a period slightly more than the \pso is known as the positive superhump period (\pop) which in turn indicates that the $\sim$ 3.7 h period in sectors 26 and 40 is \pop. The origin of \pop in CVs is thought to be due to the prograde precession of the lines of apsides of an eccentric disc \citep{1988MNRAS.232...35W, 1989PASJ...41.1005O, 1991ApJ...381..268L}, while \pom arises due to the  retrograde precession of the line of nodes of a disc that has tilted out of the orbital plane \citep{1995PASP..107..551H, 1997PASP..109.1100P, 2000ApJ...535L..39W, 2002MNRAS.335..247M}. The superorbital period found in the TESS power spectra of LS Cam could be  associated with the retrograde precession of the disc, as we have not found the precession period associated with prograde precession, which should be around $\sim$ 1.8 d. Since this scenario explains all the frequencies present in the power spectra more appropriately than that of scenario A, therefore, for further discussion, we have considered LS Cam as a superhumping CV.

\par
The positive period excess parameter $\epsilon_{+}$ with positive superhump and orbital period is related by: $\epsilon_{+}$ = (\pop - \pso)/\pso. For the negative superhump period, this relation modifies as: $\epsilon_{-}$ = (\pom - \pso)/\pso to define the negative superhump period deficit. The value of $\epsilon_{+}$ in sectors 26 and 40 was derived to be 0.090$\pm$0.002 and 0.085$\pm$0.002, respectively. We have taken an average value of 0.0875 $\pm$ 0.0035 for our further calculations. With this large value of $\epsilon_{+}$, LS Cam joins BB Dor, which has the largest value of $\epsilon_{+}$ of $\sim$0.09 \citep{2005PASP..117.1204P}.  However, the value of $\epsilon_{-}$ in sectors 19, 20, 26, and 40 was calculated to be -0.033$\pm$0.002, -0.034$\pm$0.002, -0.034$\pm$0.002, and -0.034$\pm$0.002,  and these values are well consistent with each other. Therefore, using the values of periods obtained from the combined data, $\epsilon_{-}$ was calculated to be -0.03406$\pm$0.00008. Therefore, the ratio of period excess to period deficit ($\phi$ = $\epsilon_{-}$/$\epsilon_{+}$) for LS Cam was found to be $\sim$ -0.39. The similar value for $\phi$ has also been found for TT Ari (-0.40) and TV Col (-0.36) \citep[see Table 2 of][]{2002MNRAS.330L..37R}. The superhump excess has been shown to be correlated with the mass ratio (q=M$_{2}$/M$_{1}$) of the binary system components by \citet{2005PASP..117.1204P} with the relation: $\epsilon_{+}$ = $0.18q + 0.29q^{2}$. However, this relation suffers from the difficulty of not considering pressure effect within the accretion disc as suggested by \citet{2022arXiv220102945K}. Using the above-mentioned relation of \citet{2005PASP..117.1204P}, the value of q was thus estimated to be $\sim$ 0.32. Using the mean empirical linear mass-period relation of \citet{1998MNRAS.301..767S}, $\rm M_{2}$ was estimated to be $\sim$ 0.32 M$\odot$. Using above mentioned values of q and $\rm M_{2}$, the mass of WD was estimated to be $\sim$ 1.0 M$\odot$ for LS Cam.

\begin{figure*}
\subfigure[]{\includegraphics[width=16cm, height=5cm]{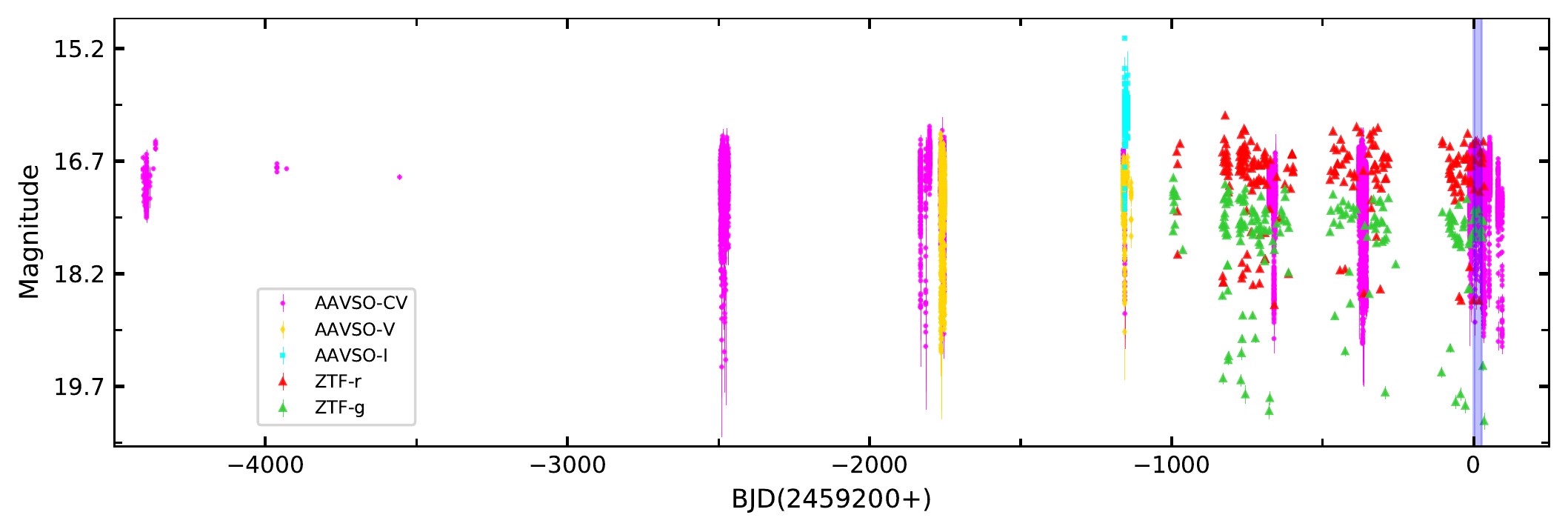}\label{fig:lc-v902-all}}
\subfigure[]{\includegraphics[width=17cm, height=10cm]{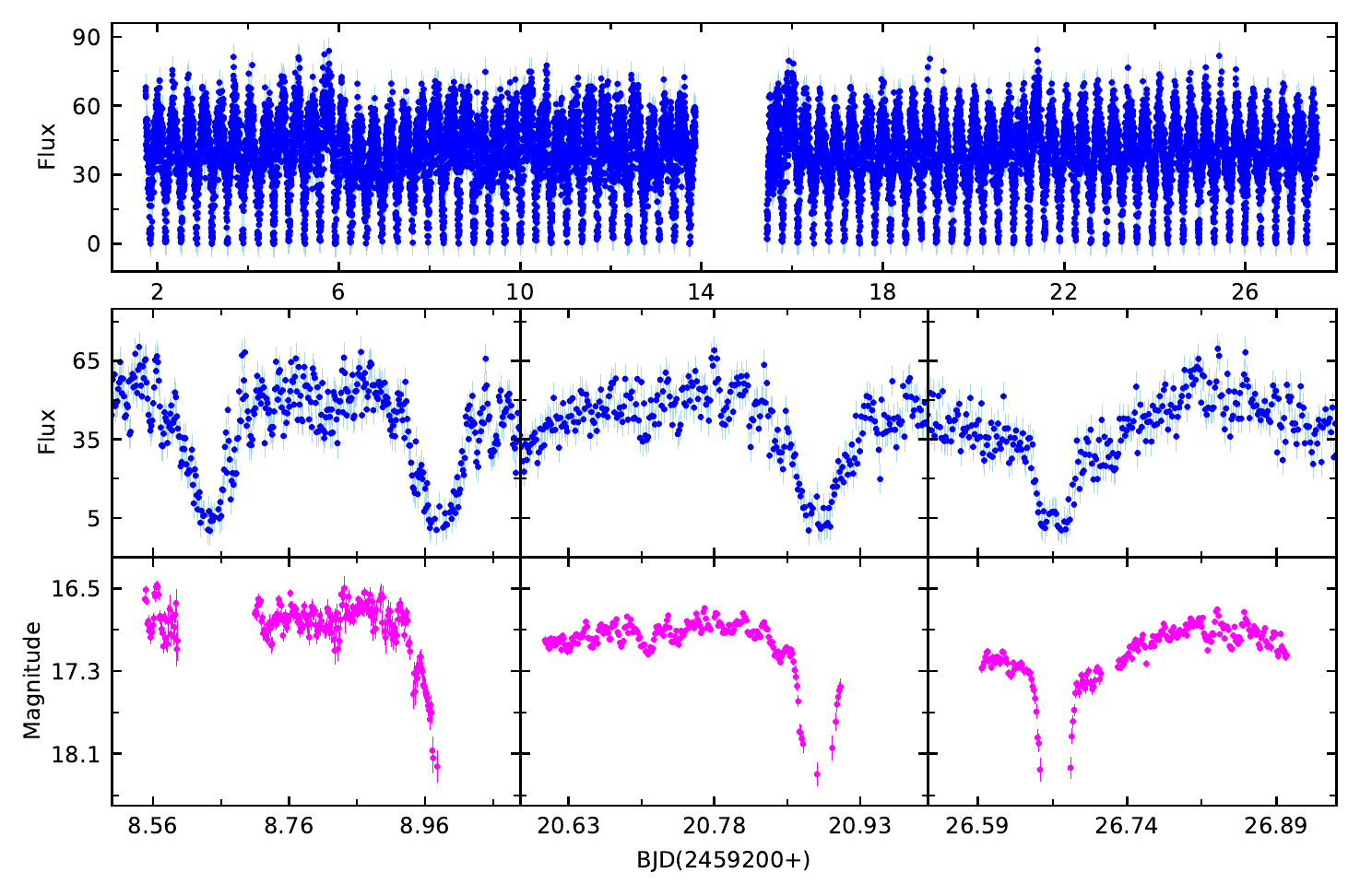} \label{fig:lc-v902}}
\caption{(a) Long-term variable light curve of V902 Mon, where the shaded region corresponds to the TESS observations. (b) Top panel: TESS light curve of V902 Mon. Middle and bottom panels show some of the simultaneous observations between TESS and AAVSO.}
\end{figure*}

\begin{figure*}
\subfigure[]{\includegraphics[width=17cm, height=5cm]{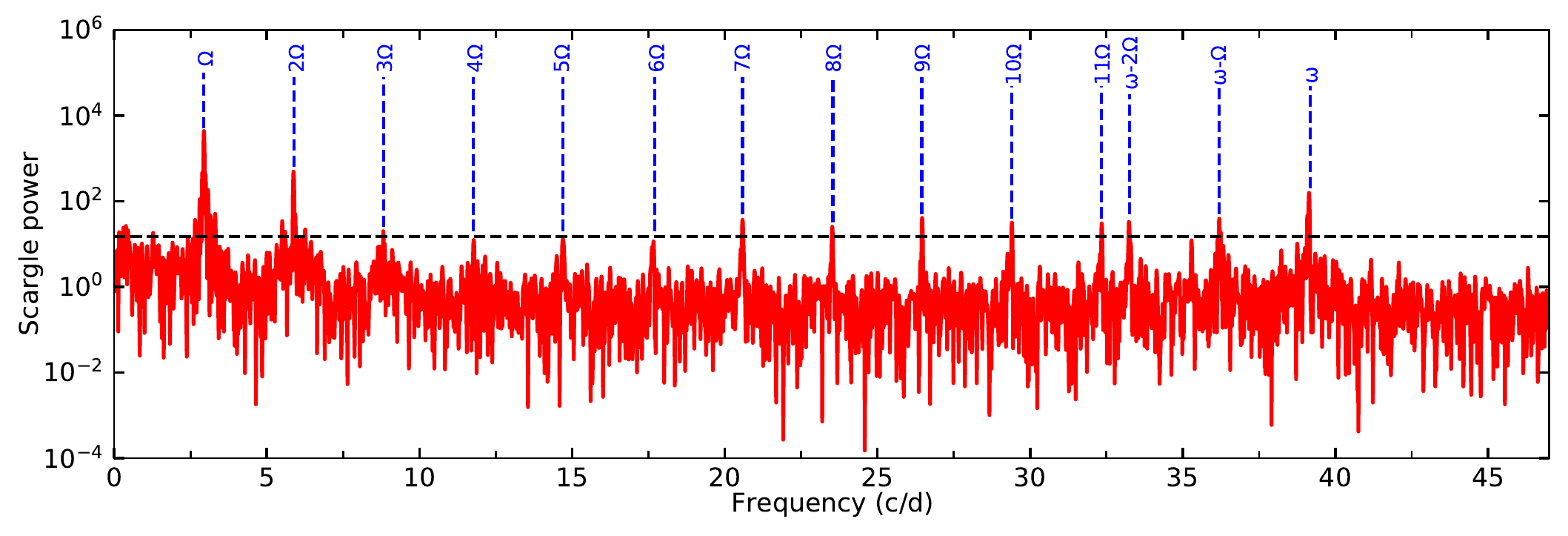}\label{fig:ps-V902Mon}}
\subfigure[]{\includegraphics[width=17cm, height=5cm]{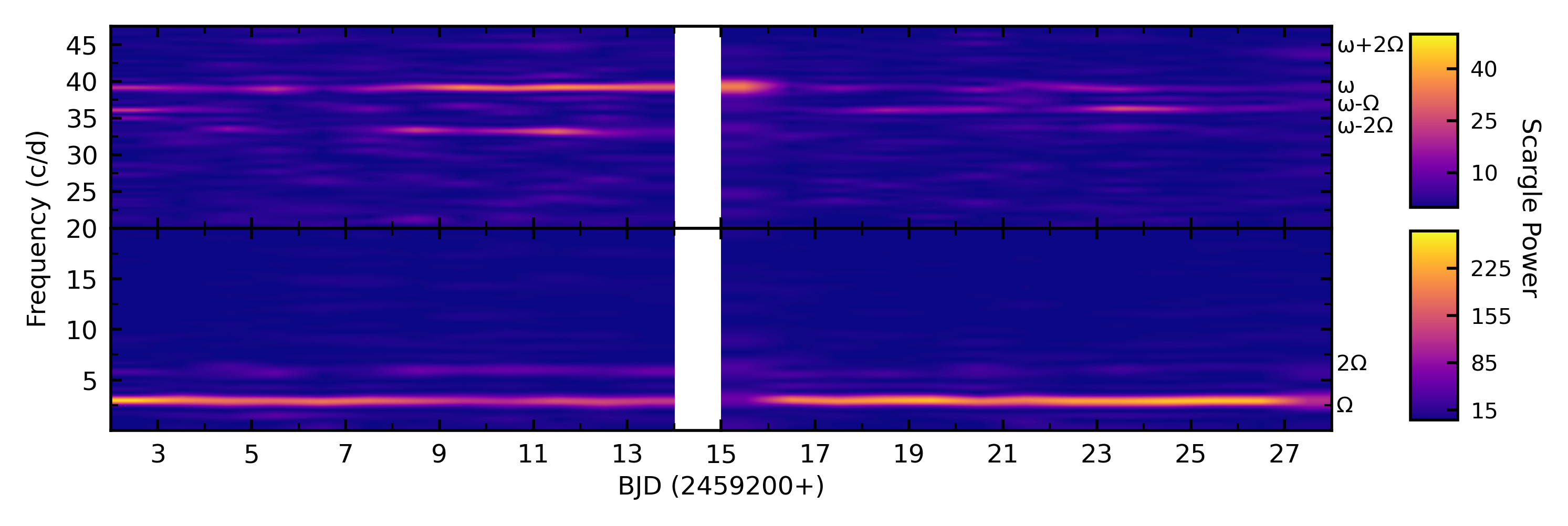}\label{fig:trail-V902Mon}}
\caption{(a) Power spectrum of V902 Mon, where the major frequencies are marked for clear visual inspection. (b) Time-resolved power spectrum of V902 Mon with a time-bin of 1 day.}
\end{figure*}

\section{V902 Mon}
\label{sec5}
\subsection{Light Curve and Power Spectral Analysis} \label{ls1}
The long-term light curve of V902 Mon is shown in Figure \ref{fig:lc-v902-all}, for which we have used data from AAVSO-CV, AAVSO-V, AAVSO-I, ZTF-r, and ZTF-g. The shaded region in the figure corresponds to the TESS observations. Figure \ref{fig:lc-v902} shows the entire TESS light curve along with some of the simultaneous data points between AAVSO and TESS. We have performed LS periodogram analysis to find the periodicities in the data, and the corresponding power spectrum is presented in Figure \ref{fig:ps-V902Mon}, where we have marked the positions of all the identified frequencies. These frequencies are $\Omega$, $2\Omega$, $3\Omega$, $4\Omega$, $5\Omega$, $6\Omega$, $7\Omega$, $8\Omega$, $9\Omega$, $10\Omega$,  $11\Omega$, \sthree, \be, and \s. All frequencies except  $4\Omega$,  $5\Omega$, and $6\Omega$ are above the 90\% significance level, which is denoted with the black dotted line in the power spectrum. The periods corresponding to all these frequencies are given in Table \ref{tab:periods-V902Mon}. We found an orbital period of 8.16 $\pm$ 0.03 h and a spin period of 2207.6 $\pm$ 0.5 s, confirming the previous results of \citet{2012ApJ...758...79A} and \cite{2018A&A...617A..52W}. In contrast to earlier epochs, a period of 2387.0 $\pm$ 0.6 s was found in the power spectra, which we assign as the beat period of the system. The power at spin frequency was found to be higher than the power at the beat frequency in the power spectrum. In addition to that, a period of 2599.7 $\pm$ 0.8 s was also present, which corresponds to the lower orbital sideband of beat frequency (\sthree).   The simultaneous presence of \sthree along with the \be frequency confirms the origin of \sthree  to be the orbital modulation of the \be  component, as pointed by \citet{1986MNRAS.219..347W}. In addition to this, the orbital modulation of the \be  component also changes the power at \s  frequency because \be $\pm$ \orb = \sthree and \s. Further, the absence of \beplus frequency also suggests that the origin of \be  can not be the orbital modulation of \s  component; otherwise, both orbital modulations of \s frequency (\be and \beplus) should have been present in the power spectrum. Although \sthree can also be originated from the modulation of \s at half the orbital period but we did not find the presence of $\omega + 2\Omega$ in the power spectrum. Therefore, the presence of \s  and \be  frequencies along with the \sthree component in the combined power spectrum obtained from the analysis of TESS observations of late 2020 show that V902 Mon accretes via a combination of disc and stream with a dominance of disc-fed accretion.

\subsection{Ephemeris and O-C Analysis}
\begin{figure*}
\centering
\includegraphics[width=15cm, height=5cm]{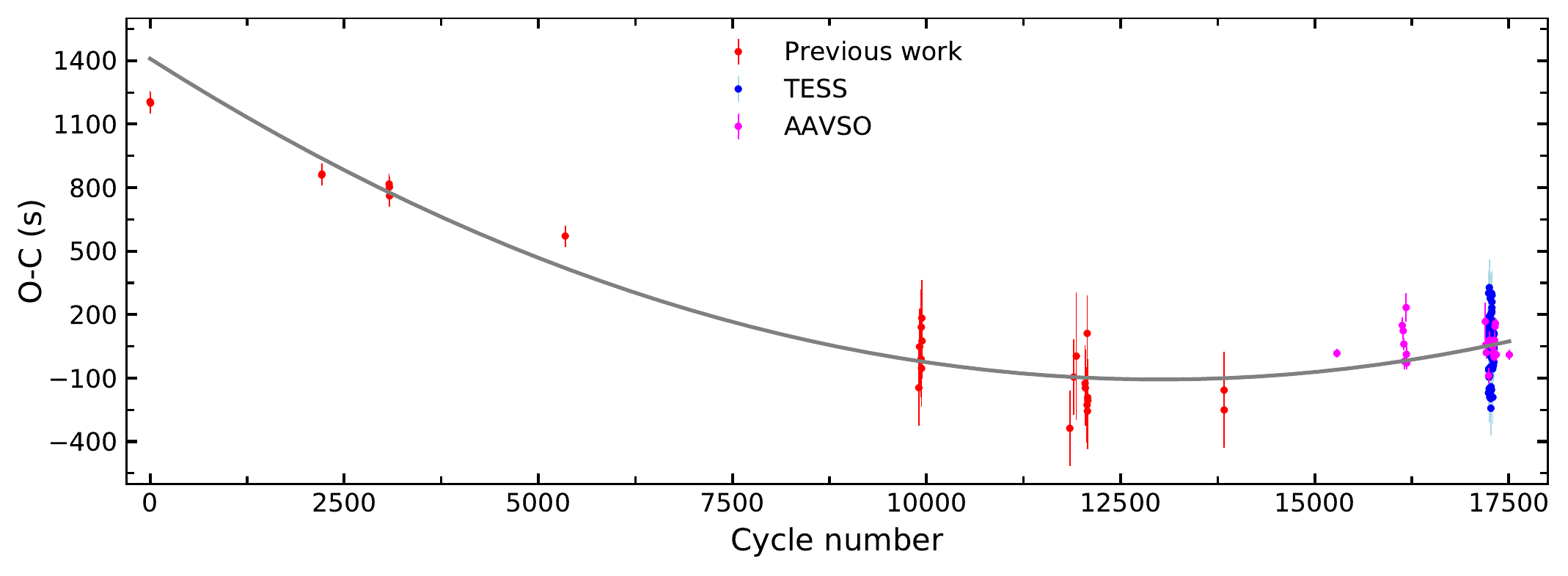}
\caption{Observed minus calculated (O-C) diagram for the source V902 Mon. The red points are from previous work done by \citet{2007MNRAS.382.1158W}, \citet{2012ApJ...758...79A}, and \citet{2018A&A...617A..52W}, whereas the blue and magenta points are from the present work as observed from TESS and AAVSO, respectively.}
\label{fig:omc}
\end{figure*}

Intending to refine the previous ephemeris, we have calculated eclipse mid-points by fitting the eclipses with a constant plus a Gaussian. We found 71 timings of minima  from TESS and 21 timings of minima from AAVSO where  5 of them were simultaneous.  The newly derived times of minima are given in Table \ref{tab:midpoint}, where errors are given in parenthesis.  We converted minima timings from AAVSO and earlier reported timings by \citet{2007MNRAS.382.1158W} and \citet{2012ApJ...758...79A} to BJD using the algorithm of \citet{2010PASP..122..935E}. We then combined the timings of the eclipse minima from our data with the previous timings of \citet{2007MNRAS.382.1158W}, \citet{2012ApJ...758...79A},  and \citet{2018A&A...617A..52W}. We have taken the uncertainties for previously reported timings as described in \citet{2018A&A...617A..52W}. A linear fit between the cycle numbers and minima timings provided the following ephemeris for V902 Mon:
\begin{equation} \label{eqn}
T_{0} = 2453340.4944(9) + 0.34008396(6) \times E   
\end{equation}
where $T_{0}$ is defined as the time of mid-eclipse and the errors are given in parenthesis.\\
From Equation \ref{eqn}, we refined the orbital period to be 8.162015 $\pm$ 0.000002 h.
Observed minus calculated (O-C) to the eclipse timings fit are shown in Figure \ref{fig:omc} and a clear trend is visible, suggesting a period change in V902 Mon. By fitting a parabola between O-C and cycle numbers, the apparent orbital period derivative was found to be (6.09 $\pm$ 0.60) $\times 10^{-10}$.

\begin{table}
\centering
\caption{Periods corresponding to the dominant peaks in the power spectrum of V902 Mon.}
\label{tab:periods-V902Mon}
\renewcommand{\arraystretch}{1.4}
\begin{tabular}{lc}
\hline
Identification & Period\\
\hline
\po (h) & 8.16 $\pm$ 0.03   \\
\ps (s) & 2207.6 $\pm$ 0.5  \\
\pb (s) & 2387.0 $\pm$ 0.6   \\
\ptwoo (h) & 4.086 $\pm$ 0.007    \\
\pthreeo (h) & 2.723 $\pm$ 0.003    \\
\pseveno (h) & 1.1662 $\pm$ 0.0006   \\
\peighto (h) & 1.0204 $\pm$ 0.0004   \\
\pnineo (h) & 0.9070 $\pm$ 0.0003   \\
\pteno (h) & 0.8162 $\pm$ 0.0003    \\
\peleveno (h) & 0.7420 $\pm$ 0.0002    \\
\psthree (s) &  2599.7 $\pm$ 0.8 \\

\hline
\end{tabular}
\end{table}

\begin{table*}
\centering
\caption{Eclipse midpoints for V902 Mon from recent observations from TESS and AAVSO.}
\label{tab:midpoint}
\renewcommand{\arraystretch}{1.4}
\begin{tabular}{lccccccc}
\hline
Eclipse midpoint (BJD) &  Cycle & Eclipse midpoint (BJD) &  Cycle & Eclipse midpoint (BJD) &  Cycle & Eclipse midpoint (BJD) &  Cycle\\
\hline

2458538.6779(2)	&	15285	&	2459205.581(1)	&	17246	&	2459213.404(1)	&	17269	&	2459222.5863(9)	&	17296	\\
2458825.0301(4)	&	16127	&	2459205.924(1)	&	17247	&	2459213.743(1)	&	17270	&	2459222.928(1)	&	17297	\\
2458829.1108(5)	&	16139	&	2459206.266(2)	&	17248	&	2459215.783(1)	&	17276	&	2459223.269(1)	&	17298	\\
2458832.1709(4)	&	16148	&	2459206.604(2)	&	17249	&	2459216.128(1)	&	17277	&	2459223.607(1)	&	17299	\\
2458834.2104(4)	&	16154	&	2459206.945(1)	&	17250	&	2459216.467(2)	&	17278	&	2459223.9473(9)	&	17300	\\
2458842.0353(8)	&	16177	&	2459207.285(1)	&	17251	&	2459216.807(1)	&	17279	&	2459224.2865(9)	&	17301	\\
2458843.0530(4)	&	16180	&	2459207.625(2)	&	17252	&	2459217.148(2)	&	17280	&	2459224.627(1)	&	17302	\\
2458844.0728(3)	&	16183	&	2459207.962(2)	&	17253	&	2459217.488(2)	&	17281	&	2459224.9689(9)	&	17303	\\
2459188.920(1)	&	17197	&	2459208.301(1)	&	17254	&	2459217.828(1)	&	17282	&	2459225.309(1)	&	17304	\\
2459190.9594(4)	&	17203	&	2459208.641(1)	&	17255	&	2459218.169(1)	&	17283	&	2459225.649(1)	&	17305	\\
2459192.9995(3)	&	17209	&	2459208.981(1)	&	17256	&	2459218.507(1)	&	17284	&	2459225.9884(8)	&	17306	\\
2459201.843(1)	&	17235	&	2459209.325(1)	&	17257	&	2459218.847(1)	&	17285	&	2459226.3272(9)	&	17307	\\
2459202.183(1)	&	17236	&	2459209.662(1)	&	17258	&	2459219.187(1)	&	17286	&	2459226.668(1)	&	17308	\\
2459202.5196(9)	&	17237	&	2459210.004(1)	&	17259	&	2459219.526(1)	&	17287	&	2459227.0081(8)	&	17309	\\
2459202.863(1)	&	17238	&	2459210.347(1)	&	17260	&	2459219.866(1)	&	17288	&	2459227.3490(9)	&	17310	\\
2459203.205(1)	&	17239	&	2459210.684(1)	&	17261	&	2459220.2058(9)	&	17289	&	2459228.7090(4)	&	17314	\\
2459203.541(1)	&	17240	&	2459211.026(1)	&	17262	&	2459220.5455(9)	&	17290	&	2459231.0903(4)	&	17321	\\
2459203.881(1)	&	17241	&	2459211.364(1)	&	17263	&	2459220.888(1)	&	17291	&	2459232.7909(3)	&	17326	\\
2459204.224(1)	&	17242	&	2459211.702(1)	&	17264	&	2459221.226(1)	&	17292	&	2459236.8702(3)	&	17338	\\
2459204.563(1)	&	17243	&	2459212.043(1)	&	17265	&	2459221.567(1)	&	17293	&	2459293.6642(3)	&	17505	\\
2459204.901(1)	&	17244	&	2459212.382(1)	&	17266	&	2459221.904(1)	&	17294	&				\\
2459205.242(1)	&	17245	&	2459212.726(1)	&	17267	&	2459222.2481(9)	&	17295	&				\\

\hline
\end{tabular}
\end{table*}

\subsection{Time Resolved Power Spectra and Phase Folded Light Curves} \label{foldv90}
We have also explored the time-resolved power spectral analysis for V902 Mon. Figure \ref{fig:trail-V902Mon} shows the power spectrum with a time binning of one day. The bottom panel of Figure \ref{fig:trail-V902Mon} shows a change in the power of the orbital frequency. However, the top panel represents the power spectrum obtained after removing the data points corresponding to the eclipse region so that the orbital modulation effect can be ignored. As visible from the figure, the one-day time-resolved power spectrum shows the change in the powers of \orb, \s, \be, and \sthree frequencies with \orb frequency being dominant. The \be frequency was not detected for 7 days in the power spectra, indicating that V902 Mon is probably accreting predominantly through the disc. For the rest 18 days, both  \s and \be  frequencies were present in the power spectra with varying dominance speculating a disc-overflow system on these days. Out of these 18 days,  the system was found to be the disc-overflow accretor with disc-fed dominance on 11 days and stream-fed dominance on 7 days. Interestingly, it was found that during days 4 and 7 to 13, \sthree was present in the power spectrum and on those days \be was not strong enough such that its orbital modulation can originate a strong power at \sthree. Therefore, the origin of \sthree during these days can be considered the modulation of \s at half the orbital period. A weak $\omega + 2\Omega$ was also found to be present during these days. The simultaneous presence of \sthree and $\omega + 2\Omega$ with unequal amplitudes was also seen in EX Hya by \citet{1989A&A...225...97S}. 
Moreover, if we compare our results with previous studies of \citet{2012ApJ...758...79A} and \citet{2018A&A...617A..52W}, we can say that those observations might have occurred during those times when accretion through the stream was not significant enough to provide a beat modulation.
\par
\begin{figure*}
\centering
\includegraphics[width=6.2cm, height=4.5cm]{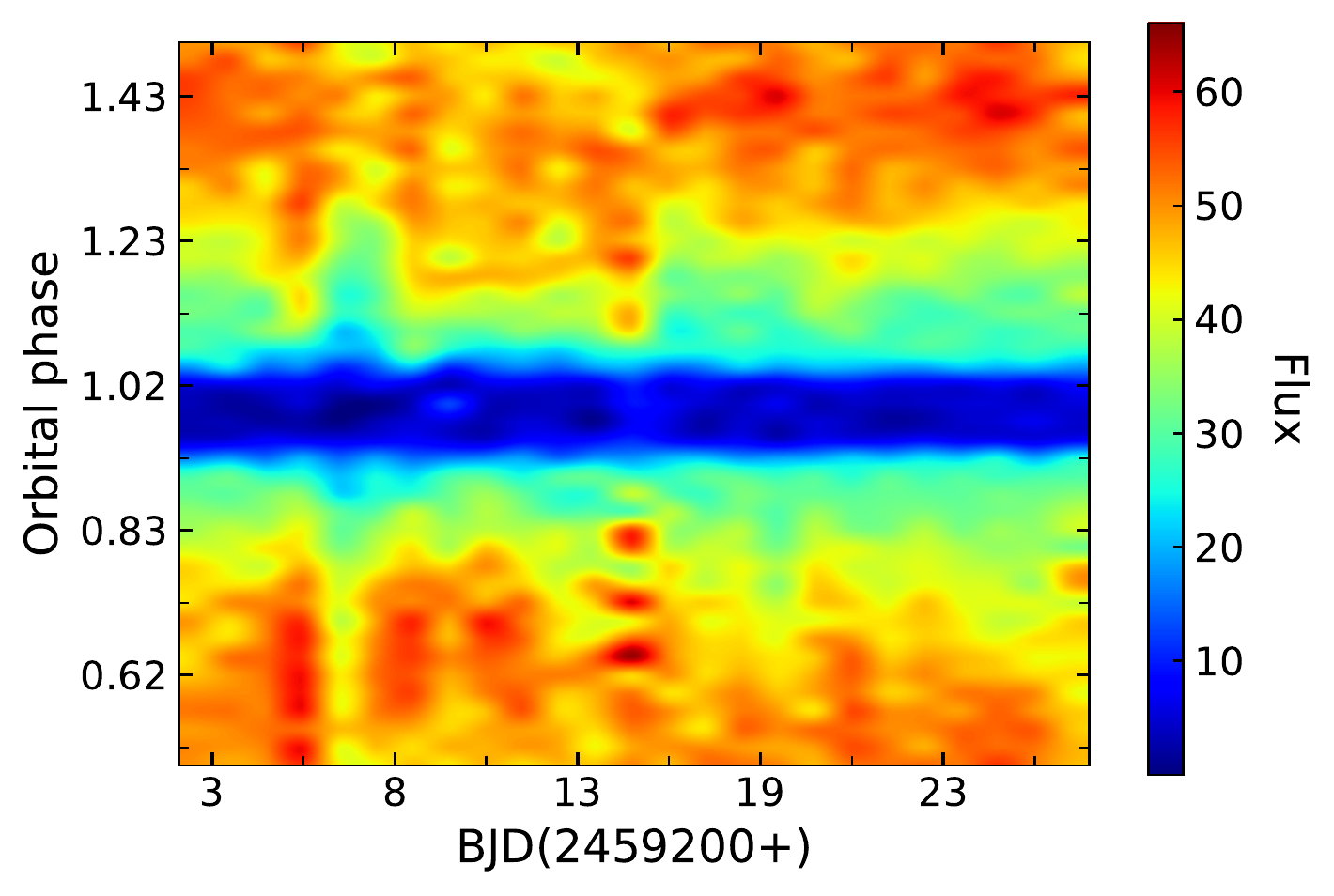}
\includegraphics[width=5.2cm, height=4.4cm]{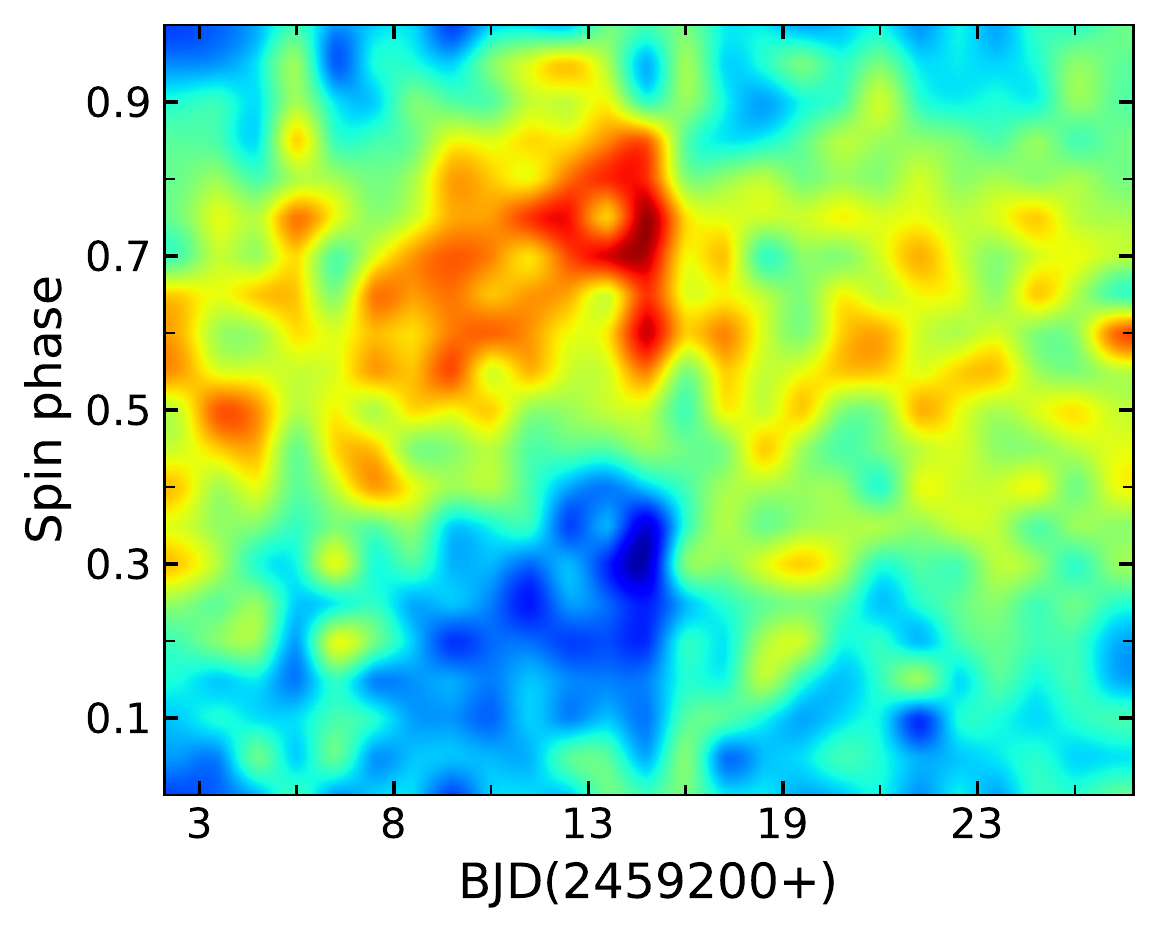}
\includegraphics[width=6.2cm, height=4.5cm]{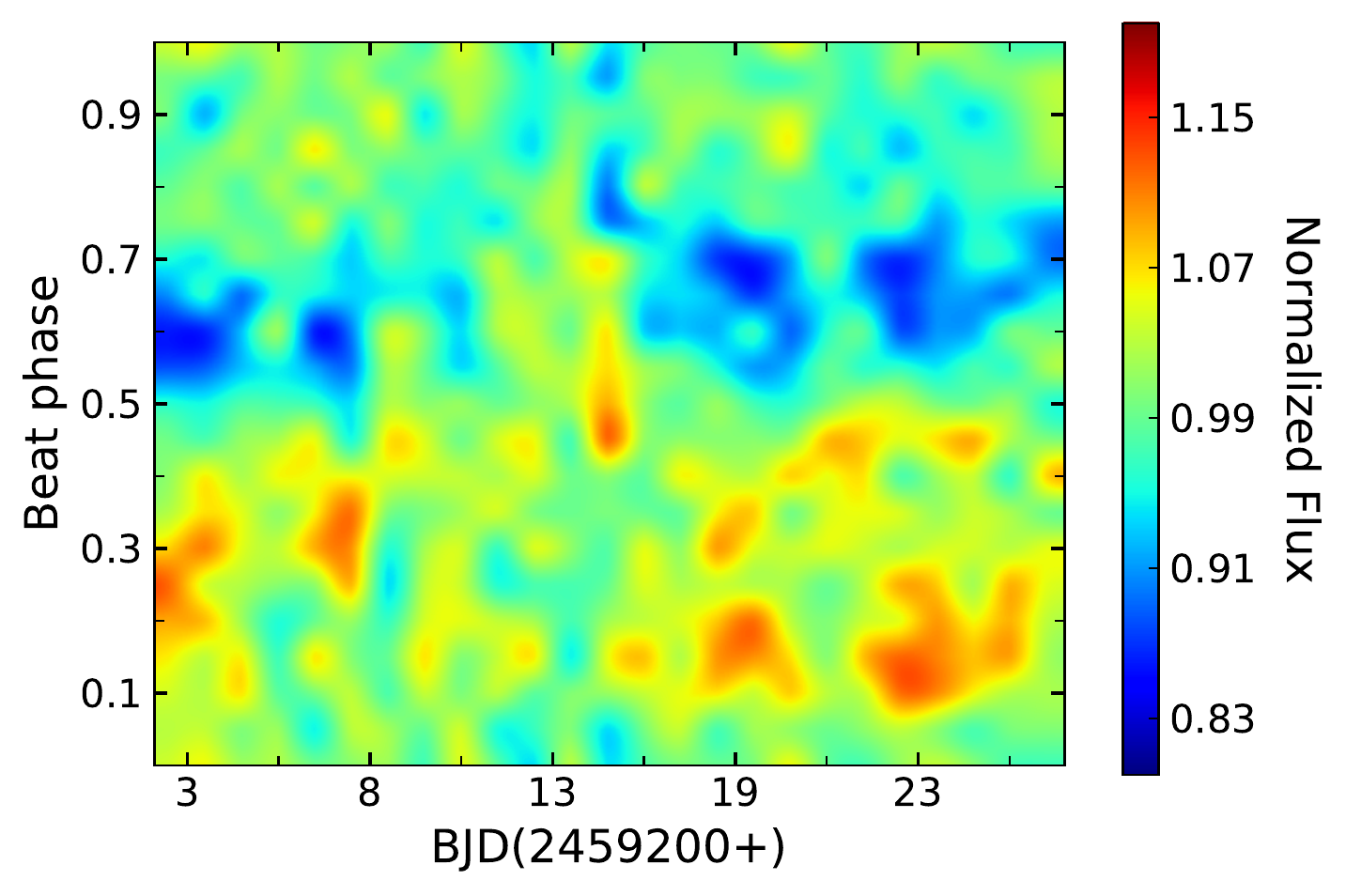}
\caption{Left panel shows orbital phase folded light curve of V902 Mon. The colourbar shows the (unnormalized) flux values. Right two panels show phase folded light curves for spin and beat periods. The same colourbar (normalized flux) has been used for representing the spin and beat modulations.}
\label{fig:phase-v902}
\end{figure*}

One-day time-resolved data segments were also taken for folding over orbital, spin, and beat periods using the updated ephemeris described in Equation \ref{eqn}. The left panel of Figure \ref{fig:phase-v902} shows the orbital phase folded light curve of V902 Mon. The out of the eclipse flux value was $\sim$ 65 e$^{-}$/s, which dropped to $\sim$ 0 e$^{-}$/s, suggesting the nature of the eclipse to be total. The eclipse phase-width was found to be almost constant during the entire observing period, which was quantified by measuring the total duration of ingress to egress in these one-day folded light curves. The region between phases 0.88 and 1.07 in the orbital folded light curve in the left panel of Figure \ref{fig:phase-v902} represents this. The data was also folded over spin and beat periods after removing the eclipse region from each dataset. 
The middle and right panels of Figure \ref{fig:phase-v902} represent the colour composite plots for spin and beat folded light curves, respectively. The interplay between the dominance of spin modulation and beat modulation is observed, which is consistent with the time-resolved power spectrum.

TESS observations of V902 Mon unveils the dynamic behaviour of its accretion geometry from pure disc-fed to variable disc-overflow. We have checked the ground-based data for V902 Mon for the epoch of TESS observations. Similar to \cite{2012ApJ...758...79A}, we have considered the out-of-eclipse brightness variations by measuring the average magnitude in the phase interval 0.78–0.88 and 1.1–1.2. It was found that during TESS observations, the out-of-eclipse magnitude of V902 Mon varies by $\sim$ 0.5 magnitudes. We have found a negative correlation of -0.8 with a null hypothesis probability of 0.027 between AAVSO magnitude and spin amplitude. We also calculated the average TESS flux in the above-mentioned phases and found a positive correlation of 0.6 with a null hypothesis probability of 0.002 between flux and spin amplitude. This suggests that during faint states, the contribution from the disc decreases, thus resulting in a lower value of spin modulation. However, we did not find a significant correlation between TESS flux and beat amplitude with a correlation value of -0.35 and a null hypothesis probability of 0.092. In previous studies by \citet{1997MNRAS.289..362N} and \citet{2020ApJ...896..116L}, a change in the mass accretion rate is typically considered the reason behind changing accretion mode in IPs. However, there are two other possibilities  given by \citet{1997MNRAS.289..362N}; the behaviour of companion star and changes in the disc itself.

\subsection{Nature of Eclipse Profiles}
To inspect whether spin pulse impacts the profile of the eclipses as seen in a previous study of \citet{2012ApJ...758...79A}, we have probed the morphology of eclipse profiles. They argued that the spin amplitude was weaker with box-shaped eclipse profiles and higher with round-shaped eclipse proﬁles. They also found that the eclipse depth was more in box-shaped profiles than the round-shaped profiles. Moreover, \citet{2018A&A...617A..52W} also found a similar result with the magnitude near the eclipse midpoint of 19.0 in flat bottom profiles rather than 18.5 in the round-shaped profile.
\par We have also explored  all these scenarios using current data. All eclipse profiles were inspected visually wherever the data were rich enough to give a clear identification. Using AAVSO-CV data, 2 flat profiles with magnitudes near eclipse midpoint of 18.6 and 18.9, and 7 round profiles with magnitudes in the range of 18.3-18.7 were found. We did not notice any clear-cut hallmark to identify the eclipse profile by measuring the mid-point brightness of the eclipse. Further, the eclipse depth for two flat profiles were 1.3 and 1.7 magnitudes, whereas, for round profiles, it has a range of 1.2-1.5 magnitudes. Therefore the high value of eclipse depth does not clearly imply that it would have flat bases. The power spectra of the  AAVSO observations corresponding to round-shaped eclipse profiles show either presence or dominance of spin frequency, whereas flat-shaped eclipse  profiles have either only beat or beat dominance.
\par In the case of TESS observations, we were able to identify 17 eclipse profiles clearly with 9 of them round-shaped and 8 of them flat-shaped.  Similar to AAVSO data, TESS observations corresponding to the round-shaped and flat-shaped eclipse profile also show either presence or dominance of spin and beat frequencies, respectively. However, the power spectra corresponding to the  one round-shaped and two flat-shaped eclipse profiles light curves have beat and spin frequency dominant power spectra, respectively.
\par We noticed that for $\sim$ 88\% of times when there is the presence of spin only or dominance of spin over beat, eclipses were round in shape, whereas for $\sim$ 89\% of times when the beat is dominant over spin, eclipses have flat bottoms.  We also suggest that the round-shaped eclipse profile corresponds to a disc dominated accretion, whereas the flat bottom eclipse profile may correspond to the stream dominated accretion.

\subsection{Orbital Inclination and Eclipsed size}
We have also estimated the inclination angle for V902 Mon using a similar approach of \citet{2012ApJ...758...79A} that incorporates the formulae given by \citet{1983ApJ...268..368E} with the known values of  the full width of the eclipse at half-depth ($\Delta \phi_{1/2}$) and mass ratio (q=M$_{2}$/M$_{1}$). From the folded light curve analysis, the mean value of $\Delta \phi_{1/2}$ is determined to be $\sim$ 0.124, which is consistent with the earlier estimates of \citet{2012ApJ...758...79A}. By using the mean empirical mass-period relation of \citet{1998MNRAS.301..767S}, M$_{2}$ falls in the range of 0.87 M$\odot$ $\lesssim$ M$_{2}$ $\lesssim$ 0.97  M$\odot$. We have used the mean WD mass value of 0.85 $\pm$ 0.21 M$\odot$ determined by \citet{2000MNRAS.314..403R} for our further calculations. Since the stable mass transfer from secondary to primary requires q $\lesssim$ 1.0, therefore for these values of M$_{1}$ and M$_{2}$, q can be estimated as 0.8 $\lesssim$ q $\lesssim$ 1.0. This finally leads to the estimation of orbital inclination of 79.4\textdegree $\lesssim$ i $\lesssim$ 83.0\textdegree, which is very well consistent with the earlier estimates by \citet{2012ApJ...758...79A} and \citet{2018A&A...617A..52W}. 

We have also determined the radius of the eclipsed region (R) by using the following equation given by \citet{1990MNRAS.243...57B}:
\begin{equation}
    R = \pi a \: \sqrt{(1-\alpha^{2})}\: \Delta \phi_{\rm ie},
\end{equation}
where a is the binary separation, $\Delta \phi_{\rm ie}$ is ingress/egress duration, and $\alpha$=cos i/cos i$_{\rm lim}$, where i is the inclination angle and i$_{\rm lim}$ is the limiting angle of inclination for which eclipse half-width at half depth reaches zero. The value of i$_{\rm lim}$ depends on mass ratio q. The average value of $\Delta \phi_{\rm ie}$ for V902 Mon is derived to be 0.047 $\pm$ 0.015, where the error
on $\Delta \phi_{\rm ie}$ is the standard deviation of different measurements. Considering the mean values of i and q, the radius of the eclipsed region is estimated to be $\sim$ 32 R$_{\rm WD}$, indicating the presence of extended emitting regions, which is also suggested by \cite{2018A&A...617A..52W}.

\begin{figure*}
\includegraphics[width=16cm, height=5cm]{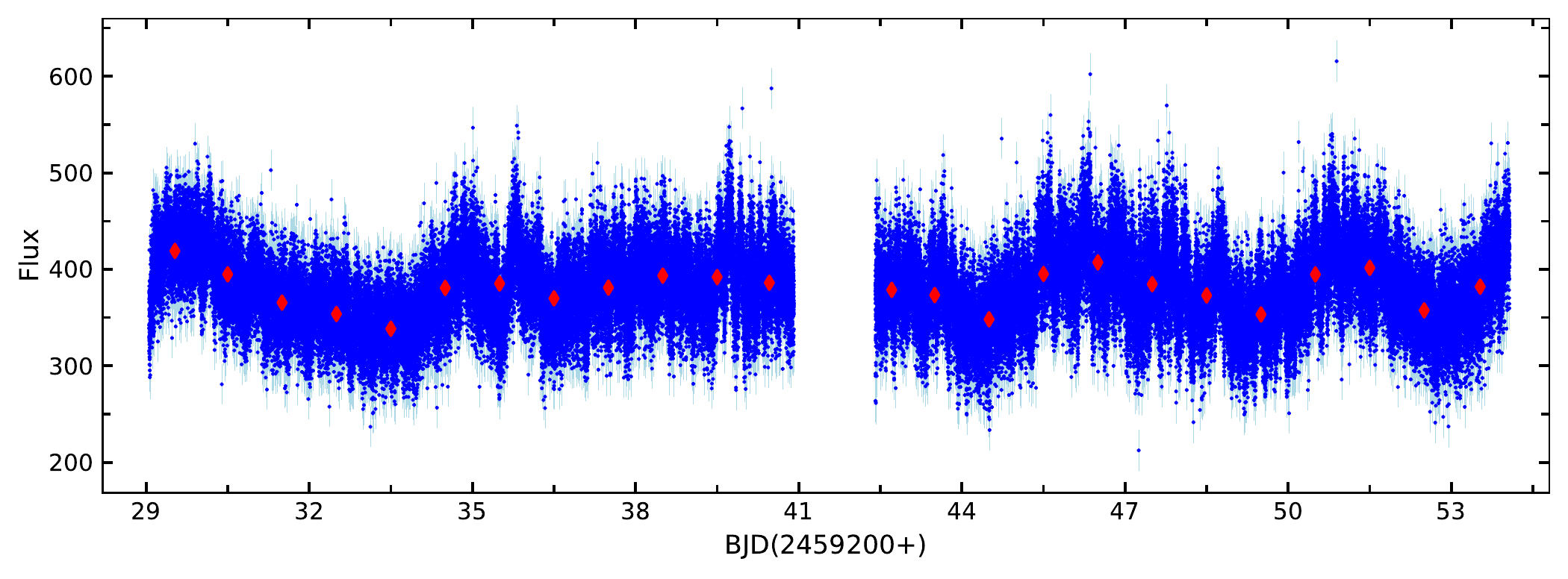}
\caption{TESS light curve of J0746, where red diamonds represent the mean flux of each day.}
\label{fig:lc-j0746}
\end{figure*}

\begin{figure*}
\subfigure[]{\includegraphics[width=17cm,height=5cm]{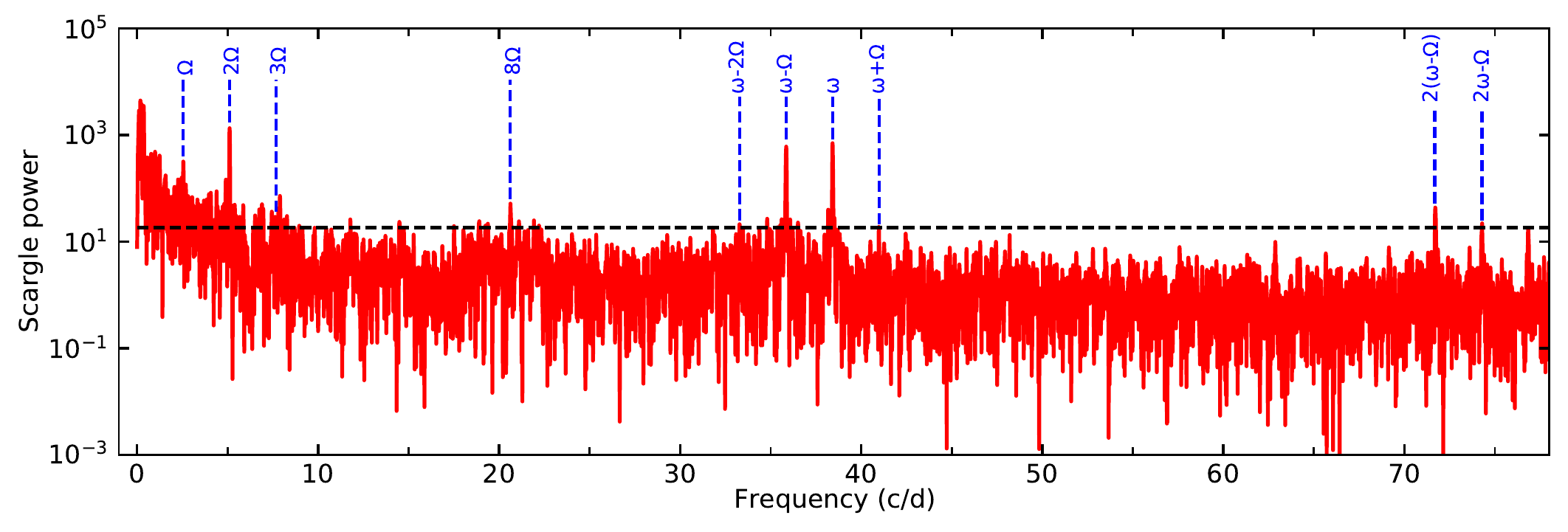}\label{fig:ps-j0746}}
\subfigure[]{\includegraphics[width=17cm, height=5cm]{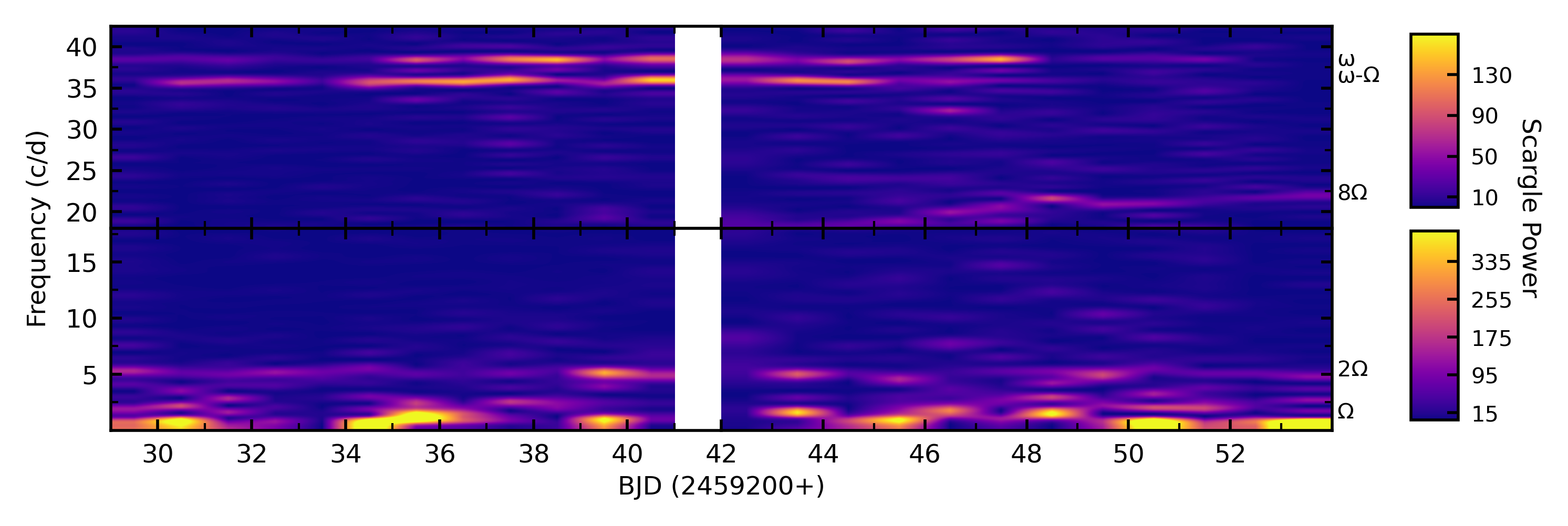}\label{fig:trail-J0746}}
\caption{(a) Power spectrum of J0746 of 20 sec cadence data. Major frequencies are marked for clear visual inspection. (b) Time-resolved power spectra of J0746 of 20 sec cadence observations with a time-bin of 1 day.}

\end{figure*}

\begin{figure*}
\includegraphics[width=5.2cm, height=4.4cm]{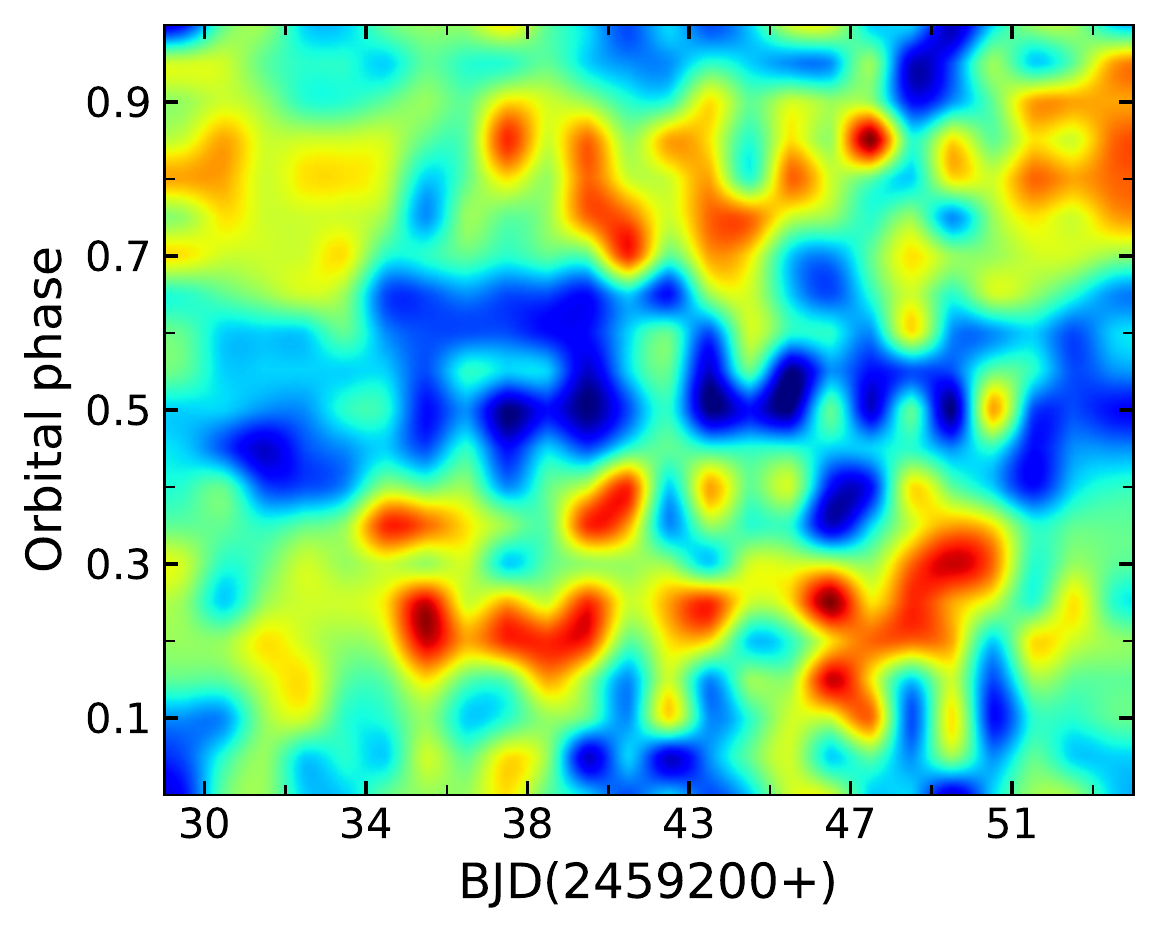}
\includegraphics[width=5.2cm, height=4.4cm]{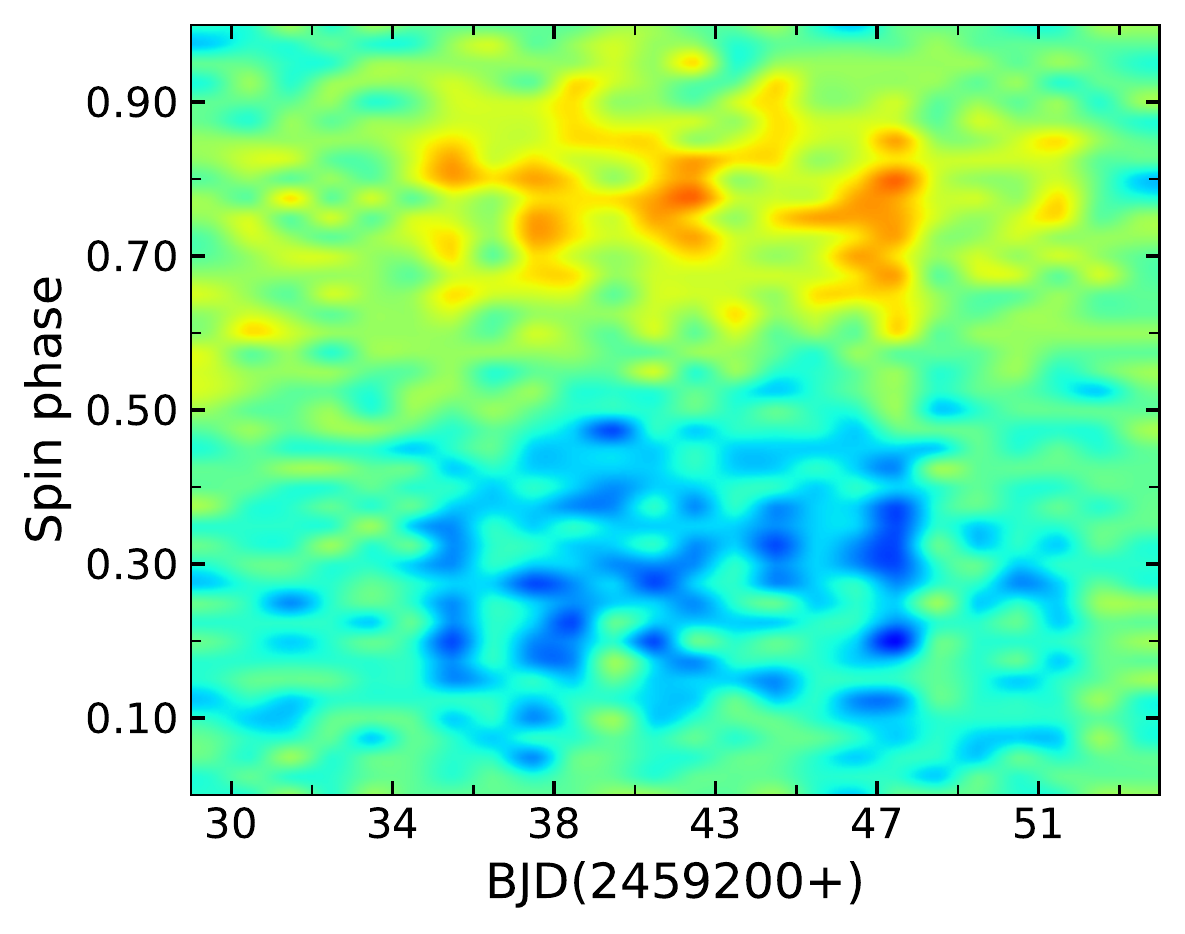}
\includegraphics[width=6.2cm, height=4.5cm]{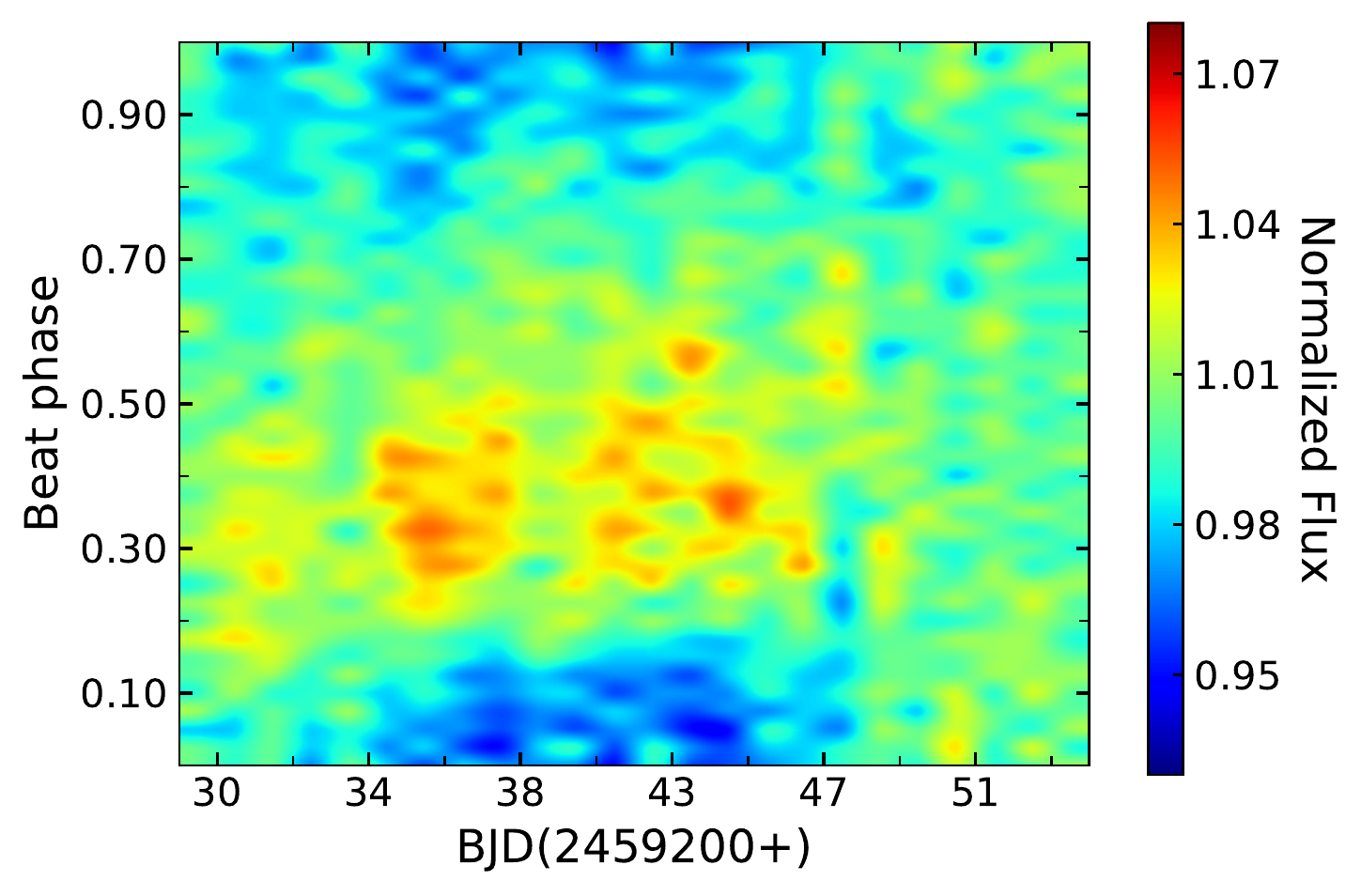}

\caption{Phase folded light curves of J0746 of 20 sec cadence observations over orbital, spin, and beat periods. The same colourbar (normalized flux) has been used for representing the orbital, spin, and beat modulations.}
\label{fig:phase-j0746}
\end{figure*}

\section{SWIFT J0746.3-1608}
\label{sec6}
\subsection{Light Curve and Power Spectral Analysis}
Figure \ref{fig:lc-j0746} represents the TESS light curve of J0746 in which  variable nature of the source is  clearly evident. Due to better time resolution with 20-sec cadence data, we have shown the LS power spectrum of the entire dataset in Figure \ref{fig:ps-j0746}. The frequencies identified in the power spectrum are \orb, \twoorb, $3\Omega$, $8\Omega$, \sthree, \be, \s,  \beplus, \twobe, and \sone. The periods corresponding to these frequencies are given in Table \ref{tab:periods-j0746}.  The orbital period obtained from the present photometric analysis of 9.38 $\pm$ 0.04 h is consistent with the earlier reported value by \cite{2013AJ....146..107T}.  However, the dominant power at \twoorb  in comparison to the \orb  frequency suggests a strong contribution from the secondary star due to its ellipsoidal modulation. Using the longer baseline available with TESS data, we have refined the spin period as 2249.0 $\pm$ 0.6 s.  The frequency corresponding to the period of  2409.5 $\pm$ 0.7 s is also  present in the power spectra and can be  attributed to the beat period of the system, which was not present in the earlier observations  \citep[see][for details]{2019MNRAS.484..101B}. 
The \s frequency was found to have slightly more power in comparison to the  \be frequency in the power spectrum indicating a disc-overflow accretion. Further, the presence of \be seems intrinsic because if \be  would have been the orbital modulation of \s  frequency, then the power at \be  and \beplus should  be almost same, but this is not the case for J0746 (see, Figure \ref{fig:ps-j0746}). These features have also been observed by \citet{1992MNRAS.254..705N} for FO Aqr, where they did not find equal power at both frequencies and suggested  that the  \be  modulation  must be intrinsic to the source. Thus we suggest the \be modulation in the source J0746  could be originating from the accretion through the stream. We speculate that the origin of \sthree  frequency can not be due to the modulation of \s  at \twoorb; otherwise, $\omega + 2\Omega$ should also be present, which was also seen in EX Hya by \citet{1989A&A...225...97S}. Therefore, we suggest that the orbital modulation of the \be component (\be $\pm$ \orb = \sthree and \s) might be the possible origin of \sthree, which can also alters the power at \s frequency. The \sone frequency, which is marginally detected, can be thought of the interaction between \s and \be frequencies because \s $\pm$ \be = \sone  and \orb. Although, in the asymmetric model of \citet{1992MNRAS.255...83W} for X-ray regimes, its presence is related to stream-fed accretion for high inclination systems. However, \citet{1999MNRAS.309..517F} have not discussed this frequency in their theoretical optical power spectral modelling. 
Furthermore a cluster of frequencies between 0.1-0.4 c/d are also present in the power spectrum with no perceivable relation to frequencies identified in J0746's power spectrum. Therefore the presence of \be, \twobe, \s, \orb, \twoorb, \sthree,  and \sone frequencies suggest that during TESS observations, J0746 seems to be accreting via a combination of disc and stream.

\begin{table}
\centering
\caption{Periods corresponding to the dominant peaks in the power spectra of 2 min and 20 sec data of J0746.}
\label{tab:periods-j0746}
\renewcommand{\arraystretch}{1.4}
\begin{tabular}{lcc}
\hline
\multirow{2}{*}{Identification} & \multicolumn{2}{c}{Period}\\
\cline{2-3}
 & 2 min & 20 sec\\ 
 \hline
\po (h) & 9.38 $\pm$ 0.04 & 9.38 $\pm$ 0.04  \\
\ps (s) & 2248.8 $\pm$ 0.6 & 2249.0 $\pm$ 0.6 \\
\pb (s) & 2409.2 $\pm$ 0.7 & 2409.5 $\pm$ 0.7   \\
\ptwoo (h) & 4.690 $\pm$ 0.009 & 4.690 $\pm$ 0.009    \\
\ptwob (s) & 1204.8 $\pm$ 0.2 & 1204.8 $\pm$ 0.2  \\
\pthreeo (h) &  ------ & 3.127 $\pm$ 0.004  \\
\peighto (h) & ----- & 1.1635 $\pm$ 0.0006 \\
\pbplus (s) &  ------ & 2108.6 $\pm$ 0.5 \\
\psthree (s) &  ------ & 2596.1 $\pm$ 0.8 \\
\psone (s) &  ------ & 1162.9 $\pm$ 0.2 \\

\hline
\end{tabular}
\end{table}

\subsection{Time Resolved Power Spectra and Phase Folded Light Curves}
Following a similar approach  as mentioned in Section \ref{ls1} for V902 Mon, one-day time-resolved power spectrum analysis has been done for J0746. The corresponding power spectrum is shown in Figure \ref{fig:trail-J0746}. All dominant frequencies \orb, \twoorb, \be, \s, and $8\Omega$ are also marked and found to have varying powers. For 22 days, both \s and \be frequencies were present in the power spectrum with varying dominance of powers between them, suggesting a disc-overflow accretion during these days. Out of these 22 days, J0746 was found to be accreting via disc-overflow with disc-fed dominance on 9 days and disc-overflow with stream-fed dominance on 13 days. For rest 2 days, both frequencies were not found to be present in the power spectra.  Further, from day 45 to 53, $8\Omega$ was found to be significant in the power spectrum.

The one-day time segments data were also taken for folding over orbital, spin, and beat periods.  The reference time for folding was the first point of observations. The phase folded light curves are shown in Figure \ref{fig:phase-j0746}. The orbital folded light curves have 2 maxima and 2 minima with a phase difference of 0.35 and 0.3, respectively. The double-humped orbital modulation seen in these orbital phase folded light curves also suggest the ellipsoidal modulation of the secondary star.  From these folded light curves, it can be easily seen the dominance of orbital modulation over the spin and beat modulation. Further, the amplitude of orbital modulation was found to be varying as the observing days progress. Moreover, from day 30 to 37, 39 to 40, 43 to 44, and 48, modulation at the beat frequency was more than the modulation at spin frequency. For days 29, 38, 42, 45 to 47, and 49 to 51, spin modulation dominates over beat modulation. The interplay between the dominance of spin and beat modulation is also consistent with the time-resolved power spectral analysis.

\par Therefore TESS observations of J0746 suggest that J0746 accretes via a combination of a disc and stream with variable dominance of both accretions. Such a change might be related to variable mass accretion rate along with the change in the activity of the secondary star, as also suggested by \citet{2019MNRAS.484..101B} for J0746.

\section{Discussion}
\label{sec7}
We have carried out detailed time-resolved timing analyses of three CVs using the high-cadence optical photometric data from TESS. From the present analyses, we speculate that LS Cam is a superhumping CV. Whereas V902 Mon and J0746 belong to the intermediate polar category of MCVs. 

\par 
We do not have strong evidence for pure IP classification of LS Cam, therefore we will discuss our results with an analogy of a superhumping CV. 
As per our knowledge, there are only 12 CVs that show simultaneous negative superhumps and superorbital periods \citep[see Table 5 of][]{2013MNRAS.435..707A} and if we consider LS Cam as a superhumping CV, then this number increases to 13. The negative superhump is `permanent' as it was found to be present in all sectors; however, the positive superhump was present only in the last two sectors of observations. The simultaneous presence of positive and negative superhumps also implies that the origin of the two can not be the same. Moreover, there are a few CVs in which both superhumps have been simultaneously detected, e.g., V503 Cyg \citep{1995PASP..107..551H}, V603 Aql \citep{1997PASP..109.1100P}, AM CVn \citep{1998ApJ...493L.105H}, V442 Oph \citep{2002PASP..114.1364P}, TT Ari \citep{2013AstL...39..111B}, and AQ Men \citep{2021MNRAS.503.4050I}. Therefore, LS Cam is an important addition to this group. The most accepted model in CVs and LMXBs to explain the simultaneous presence of superorbital period and negative superhump is the wobbling-disc model, where lines of nodes of the accretion disc precess retrogradely. This causes a negative superhump period to arise due to the interaction between precession and orbital motions. The presence of $\omega_{0}$ + N  frequency along with N frequency  could be a result of the changing of the visible disc area with the wobble frequency N \citep{2002PASP..114.1364P}.
While the origin of a tilted disc in LMXBs is known, but we do not have a clear picture for CVs. Further, a positive superhump arises due to the beating between the orbital and the prograde precession period of the elliptic accretion disc. The disc becomes elliptic because of the tidal instability which is produced due to the 3:1 tidal resonance in an accretion disc \citep{1988MNRAS.232...35W, 1989PASJ...41.1005O, 1991ApJ...381..268L}. The permanent negative superhumps have been commonly found in other kinds of CVs such as SW Sex stars, VY Scl stars, and novalike variables. Therefore, extensive X-ray and optical spectroscopic observations are required to explore the true nature of LS Cam.

\par IPs are generally clustered into three groups: the slow rotators with \ps/\po $\sim$ 0.5, intermediate with \ps/\po $\sim$ 0.1, and fast with \ps/\po $\sim$ 0.01. With \ps/\po $\sim$ 0.08 and 0.07 respectively, for V902 Mon and J0746 both fall in the category of intermediate rotators, where the majority of the IPs generally lie \citep{2004ASPC..315..216N, 2017MNRAS.470.4815B}. V902 Mon and J0746 are long orbital period systems and therefore, both are interesting from an evolutionary perspective.
V902 Mon was identified as a likely disc-accreting intermediate polar by \citet{2018A&A...617A..52W} due to the absence of \be frequency in the power spectrum. Whereas J0746 was identified as a possible IP by \citet{2019MNRAS.484..101B} and it was also found to be changing between low and high states \citep{2017MNRAS.470.4815B, 2019MNRAS.484..101B}. 
It has shown the fastest state transition in X-rays within less than a day. The detection of \be frequency for the first time along with \s frequency in the combined power spectra obtained from TESS indicate that V902 Mon and J0746 accrete via a combination of disc and stream. However, one-day time-resolved power spectra show us a bigger picture, where the power spectra seem to be changing on a timescale of days. Among disc-overflow systems,
there are two IPs, FO Aqr and TX Col, which have been
observed and studied several times.  Moreover, the change in the accretion mechanism based on the presence of spin and beat frequency in the power spectra of these IPs has been previously explored by many authors  \citep[see][for details]{1989ApJ...344..376B, 1996rftu.proc..123B, 1997MNRAS.289..362N, 2020ApJ...896..116L, 2021ApJ...912...78R}. This indicates that maybe this type of behaviour is the true nature of these systems and due to observational constraints, we were not able to detect this for the majority of IPs. Therefore, an extensive study with a larger sample is needed to connect a bridge between varying powers in the power spectra
and accretion mechanism.

\section{Summary}
\label{sec8}
We summarized our findings as:\\
\begin{itemize}
    \item [1.] A periodicity of $\sim$ 4 d along with another periodicity of 3.301 h for LS Cam in all sectors of the TESS observations is found to be present and can be attributed to the superorbital and negative superhump periods, respectively. The simultaneous presence of both periods suggests a wobbling disc model for its origin. A positive superhump period of $\sim$ 3.7 h was also found to be present in the observations of the last two sectors, which could be due to the prograde precession period of the elliptic accretion disc.  The values of period excess and period deficit were found to be 0.0875(35) and -0.03406(6), respectively. The mass ratio of the binary system components is found to be 0.321(6). \\

    \item [2.] We have detected a beat period of 2387.0(6) s for the first time in V902 Mon. Our results presented in this study hint toward the change in accretion mode during the entire period of the observation, where disc-fed dominated accretion was found to be taking place for the majority of the time. We have refined the previously confirmed spin period with a value of 2207.6(5) s as well as orbital ephemeris. Moreover, we have also found an apparent orbital period derivative of (6.09 $\pm$ 0.60) $\times 10^{-10}$. Further, we have shown that eclipsed region indicates the presence of extended emitting regions. \\

    \item [3.] In the case of J0746, a beat period of 2409.5(7) s is obtained, which was not evident in earlier studies. Moreover, using TESS observations, we were able to refine the previously reported spin period as 2249.0(6) s. Our results suggest variable accretion mechanisms taking place during the entire observation period. More than half of the observing time, J0746 was found to be stream-fed dominant accretor. The dominant modulation at \twoorb seems to be due to the ellipsoidal modulation of the secondary star.
\end{itemize}

\section{Acknowledgements}
We thank the anonymous referee for providing useful comments and suggestions that led to the significant improvement of the quality of
the paper. This paper includes data collected with the TESS mission, obtained from the MAST data archive at the Space Telescope Science Institute (STScI). Funding for the TESS mission is provided by the NASA Explorer Program. Based on observations obtained with the Samuel Oschin 48-inch and the 60-inch Telescope at the Palomar Observatory as part of the Zwicky Transient Facility project. ZTF is supported by the National Science Foundation under Grant No. AST-1440341 and AST-2034437 and a collaboration including Caltech, IPAC, the Weizmann Institute for Science, the Oskar Klein Center at Stockholm University, the University of Maryland, the University of Washington, Deutsches Elektronen-Synchrotron and Humboldt University, Los Alamos National Laboratories, the TANGO Consortium of Taiwan, the University of Wisconsin at Milwaukee, Trinity College Dublin, Lawrence Livermore National Laboratories, Lawrence Berkeley National Laboratories, and IN2P3, France. Operations are conducted by COO, IPAC, and UW.  We acknowledge with thanks the variable star observations from the AAVSO International Database contributed by observers worldwide and used in this research. NR acknowledges Mr. Prajjwal Rawat and Mr. Jaydeep Singh for technical discussion.

\section{DATA AVAILABILITY}
The data sets were derived from the TESS data archive available at \url{https://archive.stsci.edu/missions-and-data/tess}. The data underlying this article will be shared on reasonable request to the corresponding author.

\bibliographystyle{mnras}
\bibliography{3IP_tess.bib}

\appendix

\label{lastpage}
\end{document}